\documentclass[a4paper, 
10pt
]{article}

\usepackage[a4paper,left=3cm,right=3cm,top=2.5cm,bottom=2.5cm]{geometry}
\usepackage{tikz, pgfplots}
\pgfplotsset{compat=newest}

\usepgfplotslibrary{colorbrewer} 
\pgfplotsset{
    every axis/.append style={grid, width=0.75\textwidth, height=0.375\textwidth, legend cell align={left}},
    every axis plot/.append style={thick},
    every error bar/.append style={semithick, solid},
    every colorbar/.append style={font=\footnotesize},
    legend style={font=\footnotesize},
    label style={font=\footnotesize},
    tick label style={font=\footnotesize},
    cycle list/Set1-5,
    cycle multiindex* list={
        mark list*\nextlist
        Set1-5\nextlist
        },
}

\pgfplotsset{
    scatter marker/.style={draw=black, thin, mark=*, mark size=3pt}
}
\usepackage{amsmath,amssymb}
\usepackage{bm}
\usepackage{graphicx,import}
\usepackage{calc}
\usepackage{subcaption, siunitx}
\usepackage{algpseudocode,algorithm,algorithmicx}
\usepackage{subfiles}
\usepackage{authblk}
\usepackage{hyperref}

\begin{document}

\title{Hybrid neural network reduced order modelling for turbulent flows with geometric parameters}

\author[1]{Matteo~Zancanaro\footnote{matteo.zancanaro@sissa.it}}
\author[2]{Markus~Mrosek\footnote{markus.mrosek@volkswagen.de }}
\author[1]{Giovanni~Stabile\footnote{giovanni.stabile@sissa.it}}
\author[2]{Carsten~Othmer\footnote{carsten.othmer@volkswagen.de}}
\author[1]{Gianluigi~Rozza\footnote{gianluigi.rozza@sissa.it}}

\affil[1]{Mathematics Area, mathLab, SISSA, via Bonomea 265, I-34136
  Trieste, Italy}
\affil[2]{Volkswagen AG, Innovation Center Europe, 38436 Wolfsburg, Germany} 

\maketitle




\begin{abstract}
Geometrically parametrized Partial Differential Equations are nowadays widely used in many different fields as, for example, shape optimization processes or patient specific surgery studies. The focus of this work is on some advances for this topic, capable of increasing the accuracy with respect to previous approaches while relying on a high cost-benefit ratio performance. 
The main scope of this paper is the introduction of a new technique mixing up a classical Galerkin-projection approach together with a data-driven method to obtain a versatile and accurate algorithm for the resolution of geometrically parametrized incompressible turbulent Navier-Stokes problems.
The effectiveness of this procedure is demonstrated on two different test cases: a classical academic back step problem and a shape deformation Ahmed body application.
The results show into details the properties of the architecture we developed while exposing possible future perspectives for this work.
\end{abstract}

\section{Introduction}
Shape optimization in the context of turbulent flow problems is a particularly challenging task. The difficulty is linked with both the high-dimensionality of the problems that need to be solved and the number of configurations to test, the first one due to the physics, the second one due to the scope of the research. These two features make usually the problem intractable with standard numerical methods (e.g., finite element, finite volume, finite difference methods). Reduced order models \cite{handbook1,handbook2} (ROMs) are a possible tool that can be used in such a setting to make the problem solvable. There exist a variety of reduced order modeling techniques but the overall principle of all of them is to unveil a low dimensional behavior of a high dimensional system to allow faster computation. 

ROMs can be classified depending on the technique used to approximate the solution manifold and the method used to evolve the latent dynamics. The most used techniques to evaluate the solution manifold are based on linear approximation methods such as the reduced basis with a greedy approach (\cite{iapichino2014reduced, salmoiraghi2018free}), the proper orthogonal decomposition (\cite{stabile2018finite}) or non-intrusive methods as exposed in \cite{tsiolakis2020nonintrusive} but more recently also nonlinear methods have been proposed (\cite{lee2020model, kim2020efficient}). For what concerns the evolution of the latent space dynamics arguably the most common approach is based on (Petrov-) Galerkin projection of the original system onto the reduced subspace/manifold \cite{Benner2015}. Data driven techniques \cite{Brunton2019}, which are solely based on the reconstruction of the mapping between input and output quantities are also a possible approach. Recently, the latter techniques received particular attention also due to the latest discoveries in machine learning. Data-driven methods are usually easier to implement and permit to obtain efficient ROMs also in the case of nonlinear/non-affine problems and in the case of commercial codes with no access to the discretized full order system. On the other hand, they usually do not exploit information concerning the underlying physical principles and they might require a large number of training data to produce accurate results. Projection based techniques, thanks to the projection stage, incorporate in a natural way the physical knowledge but are particularly challenging to be implemented in the case of nonlinear and non-affine problems. 

In this work we propose a hybrid approach where the underlying partial differential equations are partially treated using a standard POD-Galerkin approach and partially by neural networks data-driven approaches. This choice is dictated by both practical and theoretical considerations. The practical one concerns the idea of generating an approach that could be applied to any turbulence model without the need to modify the reduced order model. In incompressible turbulent flows there exist a large number of turbulence models, used to outflank the difficulty in solving the dissipative scales, and, using a projection-based technique, would require to create a new reduced order model for each of them. Secondly, despite the large amount of theoretical work behind turbulence models, there are still a number of empirical coefficients and this makes the overall formulation less rigorous in terms of physical principles. These considerations have been used to propose a reduced order model that could be applied to any eddy viscosity turbulence model and that exploit a projection based technique for mass and momentum conservation and a data driven approach for the reconstruction of the eddy viscosity field. The model is constructed extending the work done in \cite{HijaziStabileMolaRozza2020b,GeorgakaStabileStarRozzaBluck2020} to geometrically parametrized problems \cite{stabile2020efficient} with a modification of the approach to reconstruct the eddy viscosity mapping. 


In the first part of this work we present all the technicalities related to the implementation of the previously described hybrid method: \autoref{subsec:FOM} contains the Finite Volume discretization of the incompressible Navier-Stokes equation employed for this work, \autoref{subsec:meshMotion} explains the method we selected for the motion of the mesh due to geometrical parametrization, \autoref{subsec:ROM} introduces the reduced order model while \autoref{subsec:rSimple} gives an overview on the actual algorithm used for the resolution, \autoref{subsec:eddyViscosity} treats the eddy viscosity evaluation. The second part of the paper is devoted to the presentation of the results related to two different test cases: a classical academic back step with variable slope of the step into \autoref{subsec:academicTC} and a second, more applied, one, shown into \autoref{subsec:appliedTC}, where the flow around an Ahmed body with variable slope of the rear part is resolved, both revealing good behaviours and promising results. In the end, few considerations and possible future developments for this work are present into \autoref{sec:discussion}.

\section{Models and Methods}

\subsection{The full order problem}\label{subsec:FOM}

In this work we are interested on Reynolds Averaged Navier Stokes (RANS) problems in a geometrically parametrized setting. This section is devoted to the explanation of the full order discretization employed to obtain a high fidelity solution.

The problem we want to deal with is modeled by the following equations:
\begin{equation}\label{eq:rans}
    \begin{cases}
        \displaystyle \frac{\partial \overline{\bm{u}}}{\partial t} + \bm{\nabla} \cdot (\overline{\bm{u}} \otimes \overline{\bm{u}})  = \bm{\nabla} \cdot \left[-\overline{p} \mathbf{I}+\left(\nu+\nu_t \right) \left(\bm{\nabla}\overline{\bm{u}}+\left(\bm{\nabla}\overline{\bm{u}}\right)^T\right)\right] &\textrm{ in } \Omega(\bm{\mu}) \\
        \bm{\nabla} \cdot \overline{\bm{u}} = 0  &\textrm{ in } \Omega(\bm{\mu}) \\
        \overline{\bm{u}}=g_D &\textrm{in }  \Gamma_{D} \\
        \nu \displaystyle \frac{\partial \overline{\bm{u}}}{\partial \bm{n}} - \overline{p} \bm{n} = g_N &\textrm{in } \Gamma_{N}
    \end{cases},
\end{equation}
where $\overline{\bm{u}} = \overline{\bm{u}}(t,\bm{x},\bm{\mu})$ stands for the time averaged velocity field, $\overline{p} = \overline{p}(t,\bm{x},\bm{\mu})$ stands for the mean pressure field, $\nu$ is the kinematic viscosity, $\nu_t$ is the eddy viscosity, $g_D$ is the boundary value to be assigned on Dirichlet boundaries while $g_N$ is the boundary value to be assigned on the Neumann boundaries. The vector $\bm{\mu}\in \mathcal{P} \subset \mathcal{R}^p$ is representing the vector of dimension $p$ containing the parameters of the problem that, at this stage, can be both physical or geometrical without any necessity of specification. 

From now on we will consider just steady state problems. For this reason the time derivative into the momentum equation will be neglected. Moreover we get $\overline{\bm{u}}(t,\bm{x},\bm{\mu}) = \overline{\bm{u}}(\bm{x},\bm{\mu})$, $\overline{p}(t,\bm{x},\bm{\mu}) = \overline{p}(\bm{x},\bm{\mu})$ and we will refer to them as just $\overline{\bm{u}}$ and $\overline{p}$ for sake of simplicity.

For these kind of applications, the use of Finite Volume techniques is common and reliable, even though Finite Element methods are widespread used (see \cite{donea2003finite}) and mixed techniques are available too (see \cite{busto2020pod}). To approximate the problem by the use of the Finite Volume technique, the domain $\Omega(\bm{\mu})$ has to be divided into a tessellation $\mathcal{T}(\bm{\mu}) = \{ \Omega_i (\bm{\mu}) \}_1^{N_h}$ so that every cell $\Omega_i$ is a non-convex polyhedron and $\bigcup_{i=1}^{N_h}\Omega_i{(\bm{\mu})}=\Omega(\bm{\mu})$. For sake of brevity, from now on, we will refer to $\Omega_i(\bm{\mu})$ as $\Omega_i$. 

\begin{figure}
    \centering
    \begin{tikzpicture}
    \draw (-1,-0.5) -- (0,0) -- (2,0) -- (1.5,1.5) -- (3.5,2) -- (4.5,3);
    \draw (-1.5,2.5) -- (-0.5,1.5) -- (0,0);
    \draw (-0.5,1.5) -- (1.5,1.5) -- (1.5,2.5);
    \draw (1.75,-0.5) -- (2,0) -- (4,-0.5) -- (3.5,2);
    \draw (4.5,-0.8) -- (4,-0.5) -- (4.5,0.5);
    \draw (0.7,0.7) -- (2.5,0.7);
    \fill[red!] (0.7,0.7) circle (0.1cm);
    \fill[red!] (2.5,0.7) circle (0.1cm);
    \draw[thick,->] (1.75,0.75) -- (2.5,1);
    \node at (0.7,0.4) {$\overline{\bm{u}}_i$};
    \node at (2.5,0.35) {$\overline{\bm{u}}_j$};
    \node at (-0.1,1.2) {$\Omega_i$};
    \node at (3.6,-0.2) {$\Omega_j$};
    \node at (2.6,1.3) {$\bm{S_{ij}}$};
    \end{tikzpicture}
    \caption{Scheme of the relation between two neighbor cells of the tessellation $\mathcal{T}$.}
    \label{fig:cells}
\end{figure}
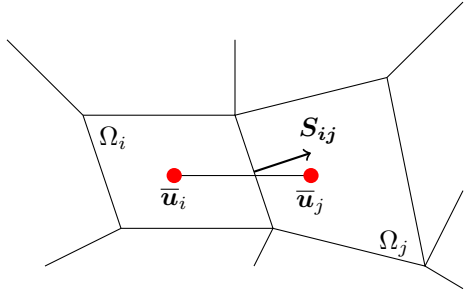

The steady-state momentum equation written in its integral form for every cell of the tessellation $\mathcal{T}$, reads as follows:

\begin{equation*}
    \int_{\Omega_i} \bm{\nabla} \cdot (\overline{\bm{u}} \otimes \overline{\bm{u}}) \, dV + \int_{\Omega_i} \bm{\nabla} \overline{p} \, dV - \int_{\Omega_i} \bm{\nabla} \cdot \left[ \left(\nu+\nu_t \right) \left(\bm{\nabla}\overline{\bm{u}}+\left(\bm{\nabla}\overline{\bm{u}}\right)^T\right)\right] \, dV =0 \ .
\end{equation*}

Let us analyze this last equation, term by term.
The convective term can be treated by the use of the Gauss' theorem:
\begin{equation*}
    \int_{\Omega_i} \bm{\nabla} \cdot (\overline{\bm{u}} \otimes \overline{\bm{u}}) \, dV = \int_{S_i} \overline{\bm{u}} \otimes \overline{\bm{u}} \cdot d\bm{S} \simeq \sum_j \bm{S}_{ij} \cdot \overline{\bm{u}}_{ij} \otimes \overline{\bm{u}}_{ij} = \sum_j \bm{F}_{ij} \overline{\bm{u}}_{ij} \ ,
\end{equation*}
where $S_i$ is the total surface related to the cell $i$, $\bm{S}_{ij}$ is the oriented surface dividing the two neighbor cells $i$ and $j$, $\overline{\bm{u}}_{ij}$ is the velocity evaluated at the center of the face $S_{ij}$ and $\bm{F}_{ij}$ is the flux of the velocity through the face $S_{ij}$ (see \autoref{fig:cells}).
Two considerations have to be underlined for this procedure. The first one is that $\overline{\bm{u}}_{ij}$ is not straight available in the sense that all the variables of the problem are evaluated at the center of the cells while here an evaluation for the velocity is required at the center of the face. Many different techniques are available to obtain it but the basic idea behind them all is that the face value is obtained by interpolating the values at the center of the cells. The second clarification is about fluxes: during an iterative process for the resolution of the equations, they are calculated by the use of the velocity obtained at previous step so that the non-linearity is easily resolved.

We now deal with the pressure term exploiting the gradient theorem:

\begin{equation*}
    \int_{\Omega_i} \bm{\nabla} \overline{p} \, dV = \int_{S_i} \overline{p} \, d\bm{S} \simeq \sum_j \bm{S}_{ij} \overline{p}_{ij} \ ,
\end{equation*}
where $p_{ij}$ is the pressure evaluated at the center of the face $S_{ij}$.

The last term to be taken into consideration is the diffusive one:
\begin{equation*}
\begin{split}
    \int_{\Omega_i} \bm{\nabla} \cdot \left[ \left(\nu+\nu_t \right) \left(\bm{\nabla}\overline{\bm{u}}+\left(\bm{\nabla}\overline{\bm{u}}\right)^T\right)\right] \, dV \simeq \left(\nu+\nu_t \right)_i \int_{\Omega_i} \bm{\nabla} \cdot  \left(\bm{\nabla}\overline{\bm{u}}+\left(\bm{\nabla}\overline{\bm{u}}\right)^T\right) \, dV \\ = \left(\nu+\nu_t \right)_i \int_{\Omega_i} \bm{\nabla} \cdot  \bm{\nabla}\overline{\bm{u}} \, dV = \left(\nu+\nu_t \right)_i \int_{S_i} \bm{\nabla}\overline{\bm{u}} \cdot d\bm{S} \simeq \sum_j \left[ \left(\nu+\nu_t \right)_{ij} \left(\bm{\nabla}\overline{\bm{u}}\right)_{ij} \right] \cdot \bm{S}_{ij} \ , 
\end{split}
\end{equation*}

where $\left(\nu+\nu_t \right)_i$ is the viscosity for the $i$-th cell, $\left(\nu+\nu_t \right)_{ij}$ is the viscosity evaluated at the center of the face $S_{ij}$ and $\left(\bm{\nabla}\overline{\bm{u}}\right)_{ij}$ refers to the gradient of the velocity evaluated at the center of the face $S_{ij}$. Notice that the gradient of the velocity is not known at the face of the cell. If the mesh is orthogonal, the approximation of its flux is straightforward:

\begin{equation*}
    \bm{S}_{ij} \cdot \left(\bm{\nabla}\overline{\bm{u}}\right)_{ij} \simeq \lvert \bm{S}_{ij} \rvert \frac{\overline{\bm{u}}_i - \overline{\bm{u}}_j}{\lvert \bm{d} \rvert} \ ,
\end{equation*}

where $\bm{d}$ is the vector connecting the centers of cells $i$ and $j$. If the mesh is not orthogonal (see \autoref{fig:cells}), a correction has to be added:

\begin{equation*}
    \bm{S}_{ij} \cdot \left(\bm{\nabla}\overline{\bm{u}}\right)_{ij} \simeq \lvert \bm{\pi}_{ij} \rvert \frac{\overline{\bm{u}}_i - \overline{\bm{u}}_j}{\lvert \bm{d} \rvert} + \bm{\omega}_{ij} \cdot \left(\bm{\nabla}\overline{\bm{u}}\right)_{ij} \ ,
\end{equation*}

where $\bm{S}_{ij}$ has been decomposed into a component parallel to $\bm{d}$, namely $\bm{\pi}_{ij}$, and another one orthogonal to $\bm{d}$, namely $\bm{\omega}_{ij}$. The term $\left(\bm{\nabla}\overline{\bm{u}}\right)_{ij}$ is finally evaluated by interpolation starting from the values $\left(\bm{\nabla}\overline{\bm{u}}\right)_{i}$ and $\left(\bm{\nabla}\overline{\bm{u}}\right)_{j}$ at the centers of the neighbor cells.

Now the complete discrete momentum equation can be written:

\begin{equation*}
    \sum_i^{N_h} \left[ \sum_j^{N_h} \bm{F}_{ij} \overline{\bm{u}}_{ij} + \sum_j^{N_h} \bm{S}_{ij} \overline{p}_{ij} - \sum_j^{N_h} \left(\nu+\nu_t \right)_{ij} \lvert \bm{\pi}_{ij} \rvert \frac{\overline{\bm{u}}_i - \overline{\bm{u}}_j}{\lvert \bm{d} \rvert} + \bm{\omega}_{ij} \cdot \left(\bm{\nabla}\overline{\bm{u}}\right)_{ij} \right] = 0 \ ,
\end{equation*}

After having applied the necessary interpolation for face centers quantities evaluation, the whole system can be rewritten into its matrix form as follow:

\begin{equation}\label{eq:discMat}
    \begin{bmatrix}
    \bm{A}_u & \bm{B}_p\\
    \bm{\nabla}(\cdot) & 0
    \end{bmatrix}
    \begin{bmatrix}
    \overline{\bm{u}}_h \\
    \overline{\bm{p}}_h
    \end{bmatrix}
    = \bm{0} \ ,
\end{equation}

where $\bm{A}_u$ is the matrix containing all the terms related to velocity into the discretized momentum equation, $\bm{B}_p$ is the matrix containing the terms related to pressure into the same equation, $\bm{\nabla}(\cdot)$ is the matrix representing the incompressibility constraint, $\overline{\bm{u}}_h$ is the vector where all the $\overline{\bm{u}}_i$ variables are collected and the same applies for $\overline{\bm{p}}_h$ with respect to $\overline{p}_i$ having $\overline{\bm{u}}_h \in \mathbb{U}_h \subset \mathcal{R}^{d \; N_h}$ and $\overline{\bm{p}}_h \in \mathbb{Q}_h \subset \mathcal{R}^{N_h}$ with $d$ spacial dimension of the problem. The interested reader can find deeper explanations on the Finite Volume discretization technique in \cite{jasak1996error, hirsch2007numerical, moukalled2016finite}.

In this work, for what concerns the offline phase, a segregated pressure-based approach has been selected. In particular, the \textit{Semi-Implicit Method for Pressure-Linked Equations} (SIMPLE) algorithm has been employed. This choice is due to the difficulties given by velocity-pressure linked problems (see e.g. \cite{caiazzo2014numerical}).

To better understand the procedure, let us report here the crucial points about this algorithm, they will be very useful later during the description of the ROM technique in this paper.

First of all we can divide the operator related to velocity into a diagonal and an extra-diagonal parts so that

\begin{equation*}
    \bm{A}_u \overline{\bm{u}}_h = \bm{A} \overline{\bm{u}}_h - \bm{H}(\overline{\bm{u}}_h)  \ .
\end{equation*}

After that, recalling \autoref{eq:discMat}, we can reshape the momentum equation as follows:

\begin{equation*}
    \bm{A} \overline{\bm{u}}_h = \bm{H}(\overline{\bm{u}}_h) - \bm{B}_p \overline{\bm{p}}_h \Rightarrow \overline{\bm{u}}_h = \bm{A}^{-1} \left[ \bm{H}(\overline{\bm{u}}_h) - \bm{B}_p \overline{\bm{p}}_h \right] \ .
\end{equation*}

In an iterative algorithm, we can express both velocity and pressure as their value at previous iteration plus a correction term:

\begin{equation*}
    \overline{\bm{u}}_h = \overline{\bm{u}}^* + \overline{\bm{u}}' \hspace{1.5cm} \overline{\bm{p}}_h = \overline{\bm{p}}^* + \overline{\bm{p}}' \ ,
\end{equation*}

where $\square^*$ terms are the old ones while $\square '$ are the corrections terms. With some approximations for the mixed terms, the following relation holds:

\begin{equation*}
    \overline{\bm{u}}_h = \bm{A}^{-1} \left[ \bm{H}(\overline{\bm{u}}^*) + \bm{H}(\overline{\bm{u}}') - \bm{B}_p \overline{\bm{p}}^* - \bm{B}_p \overline{\bm{p}}' \right] \ .
\end{equation*}

Into the SIMPLE algorithm a big assumption is taken since the extra-diagonal term $\bm{H}(\overline{\bm{u}}')$ is discarded and put to zero. Of course this makes the whole procedure no more consistent but on the counterpart it makes the resolution of the so-called \textit{pressure correction step} much easier. We then get:

\begin{equation*}\label{eq:splitA}
    \overline{\bm{u}}_h = \bm{A}^{-1} \left[ \bm{H}(\overline{\bm{u}}^*) - \bm{B}_p \overline{\bm{p}}_h \right] \ .
\end{equation*}

If we now apply the divergence operator to both sides of \autoref{eq:splitA}, we end up with a Poisson equation for pressure by exploiting the incompressibility constraint:

\begin{equation*}
    \left[ \bm{\nabla}(\cdot) \right] \overline{\bm{u}}_h =  \left[ \bm{\nabla}(\cdot) \right] \Big \{ \bm{A}^{-1} \left[ \bm{H}(\overline{\bm{u}}^*) - \bm{B}_p \overline{\bm{p}}_h \right] \Big\} \Rightarrow \left[ \bm{\nabla}(\cdot) \right] \bm{A}^{-1} \bm{B}_p \overline{\bm{p}}_h = \left[ \bm{\nabla}(\cdot) \right] \bm{A}^{-1} \bm{H}(\overline{\bm{u}}^*) \ .
\end{equation*}

\subsection{Mesh motion}\label{subsec:meshMotion}
When working in a finite volume environment, the geometrical parametrization matter is complex to be approached and treated. Some points have to be considered before starting:
\begin{itemize}
    \item as shown in \autoref{subsec:FOM}, also element-wisely, all the equation are written in their physical domain;
    \item a finite volume mesh does not have a standard cell shape, resulting on an almost random-shaped polyhedra collection;
    \item mapping the equations to a reference domain may require the use of a non-linear map but this choice wold lead to a change in the nature of the equations of the problem (see \cite{drohmann2009reduced}).
\end{itemize}
For all the reasons above, it may not be a good idea to rewrite the problem into a reference geometry to map it back to the real domain at the end of the resolution.

On the contrary in this work we decided to operate always on the real domains, moving the real mesh both during the offline and online phases. In fact, since no mapping is used, also at the online level everything is calculated in the real domain that has to be modeled according with the online parameter. This is the reason why we need a very efficient strategy for the mesh motion: in case it takes too much effort to be carried out, it compromises all the benefit coming from the reduction.

\begin{figure}
    \centering
    \begin{tikzpicture}
    \draw [line width=0.75mm] (-0.5,2) -- (2,2) -- (2,0) -- (4.5,0);
    \draw [line width=0.25mm, black!] (0,-1.5) -- (0,3.5);
    \draw [line width=0.25mm, black!] (1,-1.5) -- (1,3.5);
    \draw [line width=0.25mm, black!] (2,-1.5) -- (2,3.5);
    \draw [line width=0.25mm, black!] (3,-1.5) -- (3,3.5);
    \draw [line width=0.25mm, black!] (4,-1.5) -- (4,3.5);
    \draw [line width=0.25mm, black!] (-0.5,-1) -- (4.5,-1);
    \draw [line width=0.25mm, black!] (-0.5,0) -- (4.5,0);
    \draw [line width=0.25mm, black!] (-0.5,1) -- (4.5,1);
    \draw [line width=0.25mm, black!] (-0.5,2) -- (4.5,2);
    \draw [line width=0.25mm, black!] (-0.5,3) -- (4.5,3);
    \fill[red!] (1,2) circle (0.1cm);
    \fill[red!] (2,1) circle (0.1cm);
    \fill[red!] (3,0) circle (0.1cm);
    \fill[black!] (0,2) circle (0.1cm);
    \fill[black!] (2,2) circle (0.1cm);
    \fill[black!] (2,0) circle (0.1cm);
    \fill[black!] (4,0) circle (0.1cm);

    \draw [line width=0.75mm] (5.2,2) -- (8,2) -- (10,0) -- (12.8,0);
    \fill[red!] (7,2) circle (0.1cm);
    \fill[red!] (9,1) circle (0.1cm);
    \fill[red!] (11,0) circle (0.1cm);
    
    \draw[blue!] (7,2) circle (1.7cm);
    \draw[blue!] (9,1) circle (1.7cm);
    \draw[blue!] (11,0) circle (1.7cm);
    \draw[thick,->] (9,1) -- (10.2,2.2);
    \node at (9.5,1.7) {$r$};
    \end{tikzpicture}
    \caption{Scheme of the RBF mesh motion procedure: original mesh on the left, deformed boundary on the right where red dots are representing the control points while blue circles show the support of the function $\varphi$.}
    \label{fig:RBF}
\end{figure}
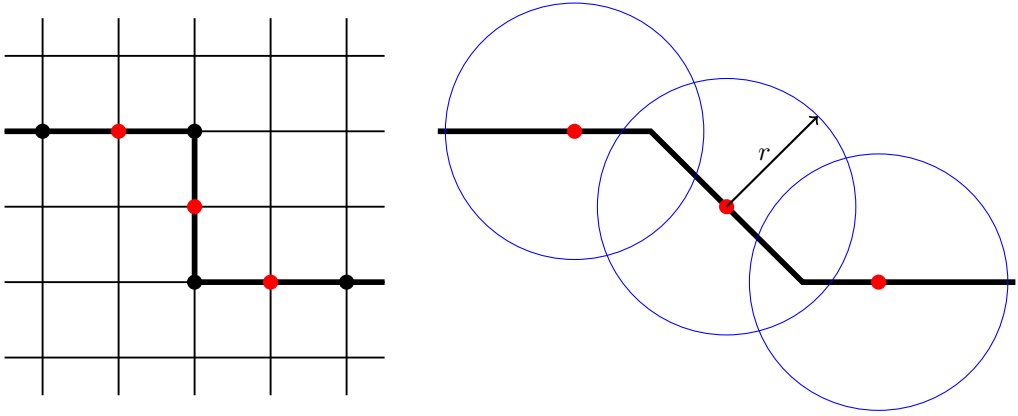

To move the mesh we use a Radial Basis Function (RBF) interpolation strategy \cite{de2007mesh}. The general formula for the evaluation of the displacements of the grid reads:

\begin{equation}\label{eq:RBF}
    \bm{\delta}(\bm{x}) = \sum_{i=0}^{N_b}\omega_i \varphi \left( \| \bm{x} - \bm{x}_i^b \| \right) + q(\bm{x}) \ ,
\end{equation}

where $\bm{\delta}(\bm{x})$ is the displacement of the grid node positioned in $\bm{x}$, $N_b$ is the number of selected control points on the moving boundary, $\omega_i$ are some calculated weights, $\varphi$ is a fixed function whose support is a round area of predetermined radius $r$, $\bm{x}_i^b$ are the coordinates of the control points and $q(\bm{x})$ is a polynomial.

The procedure can be summarized in the following steps:
\begin{enumerate}
    \item select the control points into the boundaries to be moved and shift their position obeying the fixed motion rule selected for the geometry modification, accordingly with the parameter dependent displace law: they can be either all the points into the boundary or just a fraction of their total amount if the dimension of the mesh is big enough (see \autoref{fig:RBF}), since the higher is the number of control points, the bigger (and then expensive) is the resulting RBF linear problem to be solved;
    \item calculate all the parameters for the RBF to ensure the interpolation capability of the scheme: 
    \begin{equation*}
        \begin{split}
            \bm{\delta}(\bm{x}_i^b) = \overline{\bm{\delta}}_i^b \ , \\
            \sum_{i=0}^{N_b} \omega_i q(\bm{x}_i^b) = 0 \ , \\
        \end{split}
    \end{equation*}
    resulting on the solution of the following linear problem:
    \begin{equation}\label{eq:matrixRBF}
        \begin{bmatrix}
        \bm{\Phi} & \bm{P}\\
        \bm{P^T} & 0
        \end{bmatrix}
        \begin{bmatrix}
        \bm{\omega} \\
        \bm{\alpha}
        \end{bmatrix}
        = \begin{bmatrix}
        \overline{\bm{\delta}}^b \\
        \bm{0}
        \end{bmatrix} \ ,
    \end{equation}
    where $\bm{\Phi} \in \mathcal{R}^{N_b \times N_b}$ contains the evaluations $\varphi \left( \| \bm{x}_i^b - \bm{x}_j^b \| \right)$, $\bm{P} \in \mathcal{R}^{N_b \times (d+1)}$, with spacial dimension $d$, is filled as $\left[1 \quad \bm{x}_i^b \right]$ for each row, $\bm{\alpha}$ contains the coefficients for the polynomial $q(\bm{x})$ and $\overline{\bm{\delta}}^b$ are the displacements for the control points, known a priori (see \cite{bos2013radial});
    \item evaluate all the remaining points of the grid by applying \autoref{eq:RBF}.
\end{enumerate}

Few aspects have to be underlined about the procedure above:
\begin{itemize}
    \item \autoref{eq:RBF} is used not just to move the internal points of the grid but also the points located on the moving boundaries that are not selected as control points: even if their displacement could be calculated exactly, changing their position by rigid translation while all the points of the internal mesh are shifted by the use of the RBF may lead to a corrupted grid;
    \item \autoref{eq:matrixRBF} requires the resolution of a dense linear problem whose dimension is equal to $N_b +d +1$. Thus, the number of control points have to be carefully selected. Fortunately the resolution of \autoref{eq:matrixRBF} has to be carried out just once, storing all the necessary parameters to be used in the following mesh motions;
    \item by the use of this mesh motion strategy, one ends up with meshes having all the same topology which is an important feature when different geometries have to be compared.
\end{itemize}
\subsection{The reduced order problem}\label{subsec:ROM}

The resolution of \autoref{eq:rans} for many different values of the parameter may become unaffordable. For this reason, the scope of this work, is to find an efficient way to get an accurate solution at a lower computational cost, namely a Reduced Order Model (ROM). To pursue this goal, we relay on a POD-Galerkin technique. It consists on computing a certain number of full order solutions $\bm{s}_i = \bm{s}(\mu_i)$, where $\mu_i \in \bm{T}$ for $i=1,...,N_t$, being $\bm{T}$ the training collection of a certain number $N_t$ of parameter values, to obtain the maximum amount of information from this costly stage to be employed later on for a cheaper resolution of the problem. Those snapshots can be resumed at the end of the resolution all together into a matrix $\bm{S} \in \mathcal{R}^{N_h \times N_t}$ so that:

\begin{equation}\label{eq:snapMat}
    \bm{S} = \begin{bmatrix}
    s_{1_1} & s_{2_1} & \dotsc & s_{{N_t}_1} \\
    \vdots & \vdots & \vdots & \vdots \\
    s_{1_{N^h}} & s_{2_{N^h}} & \dotsc & s_{{N_t}_{N^h}} \\
    \end{bmatrix} \ ,
\end{equation}

The idea is to perform the ROM resolution that is able to minimize the error $E_{ROM}$ between the obtained realization of the problem and its high fidelity counterpart. In the POD-Galerkin scheme, the reduced order solution can be exploited as follow:

\begin{equation*}
    \bm{s}^{ROM}(\mu) = \sum_{j=1}^{N_r} \beta_j(\mu) \bm{\xi}_j(\bm{x}) \ ,
\end{equation*}

where $N_r \leq N_{t}$ is a predefined number, namely the dimension of the reduced order solution manifold, $\beta_j(\mu)$ are some coefficients depending only on the parameter while $\bm{\xi}_j(\bm{x})$ are some precalculated orthonormal functions depending only on the position.

The best performing functions $\bm{\xi}_j$ are, in our case, the ones minimizing the $L^2$-norm error $E_{ROM}$ between all the reduced order solutions $\bm{s}_i^{ROM}$, $i=1,...,N_t$ and their high fidelity counterparts:

\begin{equation*}
    E_{ROM} = \sum_{i=0}^{N_t} \left \lVert \bm{s}_i^{ROM} - \bm{s}_i \right \rVert_{L^2} = \sum_{i=0}^{N_t} \left \lVert \sum_{j=1}^{N_r} \beta_j \bm{\xi}_j - \bm{s}_i \right \rVert_{L^2} \ .
\end{equation*}

Using a Proper Orthogonal Decomposition (POD) strategy, the required basis functions are obtained through the resolution of the following eigenproblem, obtained with the method of snapshots:

\begin{equation*}
    \bm{C} \bm{V} = \bm{V} \bm{\lambda} \ ,
\end{equation*}

where $\bm{C} \in \mathcal{R}^{N_t \times N_t}$ is the correlation matrix between all the different training solutions, $\bm{V} \in \mathcal{R}^{N_t \times N_t}$ is the matrix containing the eigenvectors and $\bm{\lambda} \in \mathcal{R}^{N_t \times N_t}$ is the matrix where eigenvalues are located on the diagonal. All the elements of $\bm{C}$ are composed by the $L^2$ inner products of all the possible couples of truth solutions $\bm{s_i}$ and $\bm{s_j}$. Of course the choice of a POD procedure for the creation of the modal basis functions is not the only possible one, see e.g. \cite{dumon2011proper}, \cite{chinesta2017model} and \cite{hesthaven2016certified}.

What may result confusing about this last computation is the fact that the $L^2$ norm is not well defined since all the realisations are obtained for different parameter values and, thus, for different domains. In this work we overtake this problem by exploiting the fact that all the meshes have the same topology. It is then possible to define a mid-configuration by the mesh motion obtained through a specific parameter $\mu_{mid}$ resulting from:

\begin{equation*}
    \mu_{mid} = \frac{1}{N_t} \sum_{i=1}^{N_t} \mu_i \ \text{for} \; \mu_i \in T \ .
\end{equation*}

In our case we use equispaced offline parameters to compose $\bm{T}$ leading to just $\mu_{mid} = \frac{\mu_1 + \mu_{N_t}}{2}$.

The correlation matrix can then be easily assembled as:

\begin{equation*}
    \bm{C}_{ij} = \bm{s}_i^T \bm{M}_{mid} \bm{s}_j \ ,
\end{equation*}
being $\bm{M}_{mid}$ the mass matrix defined for $\Omega(\mu_{mid})$.

Finally the POD basis functions are obtained as a linear combination of the training solutions as follows:

\begin{equation*}
    \bm{\xi}_i(\bm{x}) = \frac{1}{N_t} \sum_{j=1}^{N_t} \bm{V}_{ji} \bm{s}_j (\bm{x}) \ .
\end{equation*}

\begin{figure}[H]
\centering
    \begin{tikzpicture}[xscale=0.6,>=latex]
        \draw[ultra thick,->] (0.3,-1.2) -- +(0,4) node[yshift=5pt] {};
        \draw[ultra thick,->] (0.3,-1.2) -- +(217:6) node[yshift=-5pt,xshift=-5pt] {};
        \draw[ultra thick,->] (0.3,-1.2) -- +(7,0) node[xshift=6pt] {};
        
        \path[draw] (-1,0) to[out=-50,in=150] (3,-1.5);
        \path[draw] (6,0.5) to[out=150,in=-50] (2.75,1.6);
        \draw[draw] (3,-1.5) to[out=100,in=-100] (6,0.5);
        \path[draw] (2.75,1.6) to[out=100,in=-10] (-1,0);
        \shade[left color=orange!40,right color=red!70] 
          (-1,0) to[out=-50,in=150] (3,-1.5) to[out=100,in=-100] 
          (6,0.5) to[out=150,in=-50] (2.75,1.6) to[out=100,in=-10] cycle;
        \shade[left color=green!30,right color=blue!70] 
          (-1,-5) -- (3,-5) -- 
          (6,-2) -- (2.75,-2) -- cycle;
        \draw[dashed] (-1,0)--(-1,-5);
        \draw[dashed] (3,-1.5)--(3,-5);
        \draw[dashed] (6,0.5)--(6,-2);
        \draw[dashed] (2.75,-1.4)--(2.75,-2);
        
        \draw[red, line width=0.8mm] (7.2,0.5) rectangle (8.45,1.5);
        \fill[red!10] (7.2,0.5) rectangle (8.45,1.5);
        \draw[green, line width=0.8mm] (4.7,-4.3) rectangle (5.5,-3.65);
        \fill[green!10] (4.7,-4.3) rectangle (5.5,-3.65);
        \draw[blue, line width=0.8mm] (6.1,-4.3) rectangle (7.35,-3.65);
        \fill[blue!10] (6.1,-4.3) rectangle (7.35,-3.65);
        \draw[red, line width=0.8mm] (7.55,-4.65) rectangle (8.8,-3.65);
        \fill[red!10] (7.55,-4.65) rectangle (8.8,-3.65);
        \draw[blue, line width=0.8mm] (9,-4.65) rectangle (9.8,-3.65);
        \fill[blue!10] (9,-4.65) rectangle (9.8,-3.65);
        \node at (5.8,-4) {\Large $=$};
        \node at (5.1,-4) {\Large $\bm{A}_r$};
        \node at (7.8,1) {\Large $\bm{A}_h$};
        \node at (8.2,-4.2) {\Large $\bm{A}_h$};
        \node at (6.8,-4) {\Large $\bm{\Xi}^T$};
        \node at (9.4,-4.2) {\Large $\bm{\Xi}$};
        
        \node at (1.7,-4) {\huge $\mathbb{V}_r$};
        \node at (2.2,1) {\huge $\mathbb{V}_h$};
        
        \draw[ultra thick,dashed,->] (3,1) -- (7,1) node[yshift=5pt] {};
        \draw[ultra thick,dashed,->] (2.5,-4) -- (4.5,-4) node[yshift=5pt] {};
    \end{tikzpicture}
    \caption{Projection of the full order space $\mathbb{V}_h$ over the reduced one $\mathbb{V}_r$ spanned by the basis functions $\Xi$ where $\bm{A}_h$ and $\bm{A}_r$ are the full order and reduced order matrices related to the considered problem respectively.}
    \label{fig:projection}
\end{figure}
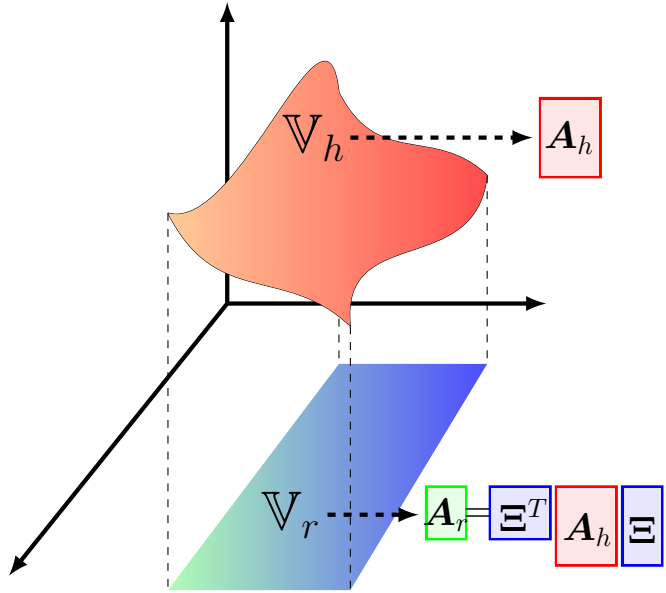

All the basis functions can be collected into a single matrix:
\begin{equation*}\label{eq:basisMat}
    \bm{\Xi} = \left[ \bm{\xi}_1, \dotsb, \bm{\xi}_{N_r} \right] \in \mathcal{R}^{N_h \times N_r} \ .
\end{equation*}
It is used to project the original problem onto the reduced subspace so that the final system dimension is just $N_r$. Supposing $N_r \ll N_h$, this procedure leads to a problem requiring a computational cost that is much lower with respect to the high fidelity one (see \autoref{fig:projection}).
Many different ways can be chosen to solve the reduced problem. For example the whole system in \autoref{eq:rans} can be assembled and projected in a monolitic approach or the equations can be treated one at a time in an iterative procedure. As we will see in \autoref{subsec:rSimple}, in this work we decided to deal with a segregated approach. This means that the momentum predictor and pressure correction steps are iterated until convergence is reached. Since the solution fields during these iterations vary a lot, from the first attempt for the variables to last resolution, the information contained into the converged snapshots is not sufficient to ensure the correct reduced reconstruction of the path to the global minimum for \autoref{eq:rans}. 

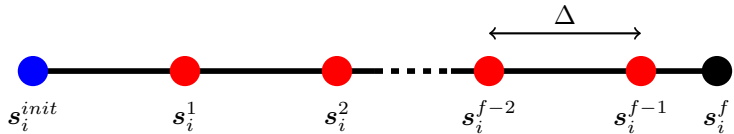
\begin{figure}[H]
    \centering
    \begin{tikzpicture}
        \draw [line width=0.75mm, black!] (5.5,0) -- (9,0);
        \draw [line width=0.75mm, black!] (0,0) -- (4.5,0);
        \draw [line width=0.75mm, black!, dashed] (4.5,0) -- (5.5,0);
    
        \fill[blue!] (0,0) circle (0.2cm);
        \fill[red!] (2,0) circle (0.2cm);
        \fill[red!] (4,0) circle (0.2cm);
        \fill[red!] (6,0) circle (0.2cm);
        \fill[red!] (8,0) circle (0.2cm);
        \fill[black!] (9,0) circle (0.2cm);
        
        \draw[thick,<->] (6,0.5) -- (8,0.5);
        \node at (0,-0.6) {$\bm{s}_i^{init}$};
        \node at (2,-0.6) {$\bm{s}_i^{1}$};
        \node at (4,-0.6) {$\bm{s}_i^{2}$};
        \node at (6,-0.6) {$\bm{s}_i^{f-2}$};
        \node at (8,-0.6) {$\bm{s}_i^{f-1}$};
        \node at (9,-0.6) {$\bm{s}_i^{f}$};
        \node at (7,0.75) {$\Delta$};
    \end{tikzpicture}
    \caption{Scheme of the snapshots selection for every parameter $\mu_i$: all red and black dots are collected together to compose the train set. Here $\bm{s}_i^{init}$ is the first attempt solution, $\bm{s}_i^j$ is the $j$-th iteration solution while $\bm{s}_i^f$ is the final converged snapshot.}
    \label{fig:snapshots}
\end{figure}

To overtake this issue, the idea proposed here is to enrich the set of snapshots for the matrix into \autoref{eq:snapMat} by the use of some intermediate snapshots that are stored during the iterations of the full order problem, as shown in \autoref{fig:snapshots}. The matrix we obtain is:
\begin{equation*}
    \bm{S} = \left[ \bm{s}_1^1, \bm{s}_1^2, \ldots, \bm{s}_1, \ldots, \bm{s}_{N_t}^1, \bm{s}_{N_t}^2, \ldots, \bm{s}_{N_t} \right] \ .
\end{equation*}
This procedure is of course somehow polluting the physical content of the resulting POD basis functions, since the intermediate steps solutions physical meaning is almost negligible, but the real gain of this procedure is to ensure a better convergence for the ROM algorithm.

\subsection{The Reduced Order SIMPLE algorithm}\label{subsec:rSimple}
We present here a new strategy for the resolution of the reduced problem: since for the full order solutions we rely on a segregated pressure based SIMPLE algorithm, the application of a monolithic approach for what concerns the online phase would lead to an inconsistency. In fact, the decoupling of the equations into the system reported in \autoref{eq:rans}, requires a slight modification of their form.
For this reason we developed a Reduced Order SIMPLE algorithm, based on the full order one, that simulates the high fidelity behaviour for what concerns the convergence to the final solution, utilizing projection-based techniques. In the following \autoref{alg:romSimple} we present the main steps for the implementation of this algorithm. For the interested reader, its laminar counterpart can be analyzed in more detail in \cite{stabile2020efficient}. Turbulence in this algorithm is treated, as it can be done for the whole SIMPLE family of algorithms, by the addition of an extra turbulent viscosity $\nu_t$ (see \cite{wilcox1998turbulence}).

Let us introduce here the snapshots matrices containing the full order solutions of \autoref{eq:rans}: $$\bm{S}_p = \left[ \overline{\bm{p}}_1, \ldots , \overline{\bm{p}}_{N_s} \right] \in \mathcal{R}^{N_h \times N_s} {,} \hspace{1cm} \bm{S}_u = \left[ \overline{\bm{u}}_1, \ldots , \overline{\bm{u}}_{N_s} \right] \mathcal{R}^{(d \; N_h) \times N_s} {,}$$ where $d$ is the space dimension of the problem and $N_s$ is the number of realizations equal to the number of provided training parameter values.

For the application of a projection-based reduction procedure of \autoref{eq:rans}, two different sets of basis functions have to be provided, for pressure and velocity respectively. This means that the procedure we exposed in \autoref{subsec:ROM} has to be carried out for both $\bm{S}_p$ and $\bm{S}_u$. Reduced pressure $\overline{p}_r$ and reduced velocity $\overline{\bm{u}}_r$ can then be written as: $$\overline{p}_r = \sum_{i=0}^{N_p} b_i \theta_i = \Theta^T \bm{b} \ ,$$ $$\overline{\bm{u}}_r = \sum_{i=0}^{N_u} a_i \psi_i = \Psi^T \bm{a} \ ,$$
where $N_p \leq N_s$ and $N_u \leq N_s$ are the selected number of modal basis functions chosen to reconstruct pressure and velocity manifolds $\mathbb{V}_p$ and $\mathbb{V}_u$ respectively, so that $\overline{p}_r \in \mathbb{V}_p = span \{\theta_1, \ldots \theta_{N_p} \}$ and $\overline{\bm{u}}_r \in \mathbb{V}_u = span \{\psi_1, \ldots \psi_{N_u} \}$, being $\theta_i$ the POD basis for pressure and $\psi_i$ the POD basis for velocity. Matrices $\Theta$ and $\Psi$ contain the modal basis functions for pressure and velocity. 

\begin{algorithm}
\caption{The Reduced Order SIMPLE algorithm}
\label{alg:romSimple}
\hspace*{\algorithmicindent} \textbf{Input:} first attempt reduced pressure and velocity coefficients $\bm{b}^\star$ and $\bm{a}^\star$; modal basis functions matrices for pressure and velocity $\Theta$ and $\Psi$\\
\hspace*{\algorithmicindent} \textbf{Output:} reduced pressure and velocity fields $\overline{p}_r$ and $\overline{\bm{u}}_r$
\begin{algorithmic}[1]
  \State From $\bm{b}^\star$ and $\bm{a}^\star$, reconstruct reduced fields $\overline{p}^\star$ and $\overline{\bm{u}}^\star$: $$\overline{p}^\star = \Theta^T \bm{b}^\star {,} \hspace{1.5cm} \overline{\bm{u}}^\star = \Psi^T \bm{a}^\star;$$
  \State Evaluate the eddy viscosity field $\nu_t$;
  \State Momentum predictor step : assemble the momentum equation, project and solve it to obtain a new reduced velocity coefficients $\bm{a}^{\star\star}$:
  	$$(\bm{\psi}_i,\bm{A} \overline{\bm{u}}^\star  - \bm{H}(\overline{\bm{u}}^\star) + \bm{\nabla} \overline{p}^\star)_{L^2(\Omega)}=0; $$
  \State Reconstruct the new reduced velocity $\overline{\bm{u}}^{\star\star}$ and calculate the off-diagonal component $\bm{H}(\overline{\bm{u}}^{\star\star})$;
  \State Pressure correction step: project pressure equation to get new reduced pressure coefficients $\bm{b}^{\star\star}$:
  		$$(\theta_i,\bm{\nabla} \cdot [ \bm{A}^{-1} \bm{\nabla} \overline{p} ] - \bm{\nabla} \cdot [\bm{A}^{-1} \bm{H}(\overline{\bm{u}}^{\star\star})])_{L^2(\Omega)}=0;$$
		Then correct the velocity explicitly after having reconstructed the new pressure $\overline{p}^{\star\star}$;
  \State Relax the pressure field and the velocity equation with the prescribed under-relaxation factors $\alpha_p$ and $\alpha_u$, respectively. The under-relaxed fields are called $\overline{p}^{ur}$ and $\overline{\bm{u}}^{ur}$;
  \If {convergence}
  \State $\overline{\bm{u}}_r = \overline{\bm{u}}^{ur}$ and $\overline{p}^\star = \overline{p}^{ur};$ 
  \Else 
  \State Assemble the conservative face fluxes $F_{ij}$:
	$$F_{ij} = \overline{\bm{u}}_{ij} \cdot \bm S_{ij};$$
  \State set $\overline{\bm{u}}^{\star} = \overline{\bm{u}}^{ur}$ and $\overline{p}^\star = \overline{p}^{ur};$ 
  \State iterate from step 1. 
  \EndIf
\end{algorithmic}
\end{algorithm}

Fluid flows projection based ROMs usually require to be stabilized in some way (see e.g. \cite{bergmann2009enablers, iollo2000stability, azaiez2021cure}). For Navier-Stokes problems, in particular, the use of stable snapshots does not guarantee the Ladyzhenskaya-Brezzi-Babushka condition fulfillment for the saddle-point problem (see \cite{ballarin2015supremizer}). The accuracy in the pressure field is of high relevance for many different configurations (see \cite{stabile2017advances}). In this case, the application of a segregated approach, also at the reduced level, leads to the complete unnecessity of extra stabilization.

Into step number $2$ of \autoref{alg:romSimple} no explanation is provided on how to evaluate the eddy viscosity $\nu_t$. This is a crucial point of the whole procedure and requires a deeper analysis that we provide to the reader in \autoref{subsec:eddyViscosity}.

\subsection{Neural Network eddy viscosity evaluation}\label{subsec:eddyViscosity}

Different possibilities are available for the closure of turbulent problems (see \cite{wang2012proper}); to make the ROM independent from the chosen turbulence model in the FOM, different approaches are eligible (see, e.g., \cite{hijazi2020effort, georgaka2020hybrid}). In this case a data-driven approach is employed for the eddy viscosity $\nu_t$. Analogously as for velocity and pressure, first, the reduced eddy viscosity $\overline{\nu_t}_r$ is computed via POD on the snapshot matrix $\bm{S}_{\nu_t} \in \mathcal{R}^{N_h \times N_s}$:

\begin{equation}
    \overline{\nu_t}_r = \sum_{i=0}^{N_{\nu_t}} c_i \zeta_i = Z^T \bm{c},
\end{equation}

where $\zeta_i$ and $c_i$ are the POD modes and coefficients for eddy viscosity, respectively, and $N_{\nu_t} \leq N_s$ denotes the selected number of modes to reconstruct the eddy viscosity. 

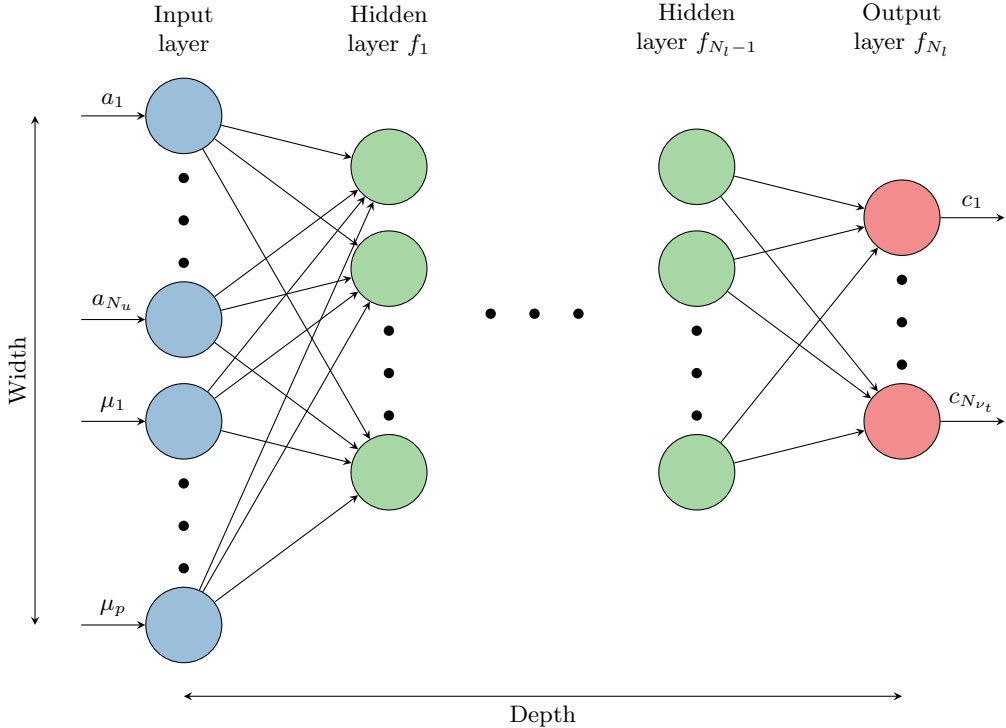
\begin{figure}[htbp]
    \centering
    \tikzset{%
      every neuron input/.style={
        circle,
        draw,
        minimum size=1cm,
        fill=Set1-B!50
      },
       every neuron hidden/.style={
        circle,
        draw,
        minimum size=1cm,
        fill=Set1-C!50
      },
       every neuron output/.style={
        circle,
        draw,
        minimum size=1cm,
        fill=Set1-A!50
      },
      neuron missing/.style={
        draw=none, 
        fill=none,
        scale=4,
        text height=0.333cm,
        execute at begin node=\color{black}$\vdots$
      },
       layer missing/.style={
        draw=none, 
        scale=4,
        text height=0.333cm,
        execute at begin node=\color{black}$\dots$
      },
    }

    \begin{tikzpicture}[x=1.5cm, y=1.5cm, >=stealth, font=\small, scale=0.9]
    
    \foreach \m/\l [count=\y] in {1, missing, 2, 3, missing, 4}
      \node [every neuron input/.try, neuron \m/.try] (input-\m) at (0,2.5-\y) {};
    
    \foreach \m [count=\y] in {1, 2, missing,3}
      \node [every neuron hidden/.try, neuron \m/.try ] (hidden1-\m) at (2,2-\y) {};
     
    \foreach \m [count=\y] in {1, 2, missing,3}
      \node [every neuron hidden/.try, neuron \m/.try ] (hidden2-\m) at (5,2-\y) {};
    
    \foreach \m [count=\y] in {1,missing,2}
      \node [every neuron output/.try, neuron \m/.try ] (output-\m) at (7,1.5-\y) {};
    

    \draw [<-] (input-1) -- ++(-1,0)
        node [above, midway] {$a_1$};
    \draw [<-] (input-2) -- ++(-1,0)
        node [above, midway] {$a_{N_{u}}$};
    \draw [<-] (input-3) -- ++(-1,0)
        node [above, midway] {$\mu_1$};
    \draw [<-] (input-4) -- ++(-1,0)
        node [above, midway] {$\mu_{p}$};
    
    
    \foreach \l [count=\i] in {1,N_{\nu_t}}
      \draw [->] (output-\i) -- ++(1,0)
        node [above, midway] {$c_{\l}$};
    
    \foreach \i in {1,...,4}
      \foreach \j in {1,...,3}
        \draw [->] (input-\i) -- (hidden1-\j);
    
    \foreach \i in {1,...,3}
      \foreach \j in {1,...,2}
        \draw [->] (hidden2-\i) -- (output-\j);
    
    \foreach \l [count=\x from 0] in {Input \\ layer, Hidden \\ layer  $f_{1}$}
    \node [align=center, above] at (\x*2,2) {\l};
    \node [align=center, above] at (5,2) {Hidden \\ layer $f_{N_{l}-1}$};
    \node [align=center, above] at (7,2) {Output \\ layer $f_{N_l}$};
    
    \node[layer missing] at (3.5, 0) {};
    
    \draw[<->] (0, -4.2) -- node[below] {Depth} (7, -4.2);
    \draw[<->] (-1.45, -3.5) -- node[above, rotate=90] {Width} (-1.45, 1.5);
      
    \end{tikzpicture}
    \caption{Illustration of a neural network that maps the POD coefficients for velocity $\bm{a} \in \mathcal{R}^{N_u}$ and the parameter values $\bm{\mu} \in \mathcal{R}^{p}$ as inputs to the the POD coefficients $\bm{c} \in \mathcal{R}^{N_{\nu_t}}$ of the eddy viscosity $\nu_t$ via $N_l$ fully connected layers.}
    \label{fig:nn_nut}
\end{figure}

In contrast to the POD coefficients of velocity and pressure, which are obtained by projecting the full order problem onto the respective POD modes and subsequently solving the reduced order problem, the POD coefficients for the eddy viscosity are modeled via a multilayer feedforward neural network. This neural network takes as the input the POD coefficients for velocity $\bm{a}$ and the corresponding geometrical parameters values $\bm{\mu}$ and maps them to the POD coefficients of the turbulent viscosity $\mathbf{\Tilde{c}}$ (Tilde denotes a prediction from the neural network) as the output (\autoref{fig:nn_nut}). 

Subsequently, the basics of multilayer feedforward neural networks and their training process are briefly reviewed; for a comprehensive description, we refer to Goodfellow et al.\ \cite{goodfellow2016}. The input to the neural network is commonly denoted as $\bm{x}$ and for our application reads:
\begin{equation}
    \mathbf{x} = \begin{pmatrix}
                    a_1 \\
                    \vdots \\
                    a_{N_u}  \\
                    \mu_1 \\
                    \vdots \\
                    \mu_p 
                  \end{pmatrix} \in \mathcal{R}^{(N_u+p)}.
\end{equation}
The choice on what to use for the input is supported by the fact that the dependency of the eddy viscosity field on the velocity field is well known because of the way the RANS equations are constructed while the dependency on the geometric parameters help in the accuracy of the network. The mapping from this input vector to the coefficients for the eddy viscosity $\mathbf{\Tilde{c}}$ is learned by the multilayer neural network via $N_l$ fully connected layers:
\begin{equation}
    \mathbf{\Tilde{c}} = f_{N_{l}}(f_{N_{l}-1}(\dots f_1(\mathbf{W_1}\mathbf{x} + \mathbf{b_1})\dots)),
    \label{eq:func_comp}
\end{equation}
where layer $i$ ($i=1, \dots, N_l$) performs an affine transformation of its input (specified by the trainable weight matrix $W_i$ and bias $b_i$) that is subsequently passed through the (linear or nonlinear) element-wise activation function $f_i$.

To train the weights $\bm{\theta} = \{\mathbf{W}_i, \mathbf{b}_i\}_{i=1}^{N_l}$ in supervised learning, the empirical risk over the training data $J$ is minimized:
\begin{equation}
    J(\bm{\theta}) = \mathbb{E}_{x \sim \Tilde{p}_{data}} \left[\mathcal{L}(\mathbf{\Tilde{c}}, \mathbf{c})\right] = \frac{1}{n_{train}} \sum_{i=1}^{n_{train}} \mathcal{L}(\mathbf{\Tilde{c}}^{(i)}, \mathbf{c}^{(i)}),
    \label{eq:nn_risk}
\end{equation}
where $\Tilde{p}_{data}$ and $n_{train}$ denote the empirical distribution of the training data and the number of training samples, respectively; $\mathcal{L}(\mathbf{\Tilde{c}}, \mathbf{c})$ is a per-sample loss metric that describes the discrepancy between target output $\mathbf{c}$ (given by training data) and predicted output $\mathbf{\Tilde{c}}$ (by neural network).

As loss function, we use the squared $L^2$-loss function (also known as mean squared error), the most common choice for the loss function in regression problems:
\begin{equation}
    \label{eq:l2_loss}
    \mathcal{L} = \|\mathbf{c} - \mathbf{\Tilde{c}} \|^2_2.
\end{equation}
Employing this loss function, the objective function $J$ is minimized using the Adam \cite{kingma2014} optimizer with minibatching, and the required gradients of the parameters with respect to the loss function are calculated via backpropagation \cite{rumelhart1986}.

The hyperparameters of the neural network, which are the parameters that are not subject to the optimization during training, were tuned for each test case separately by minimizing the loss on a designated validation data set (while the accuracy evaluation of the neural network was finally performed on a third set, referred to as test set). The hyperparameters subject to tuning were: the height and width of the neural network (i.e. the number of hidden layers and units per hidden layer, cf. \autoref{fig:nn_nut}), the activation functions for each layer, and the learning rate as well as the batch size of the Adam optimizer. For the creation and training of the neural networks, we employed the Python library PyTorch \cite{paszke2019}.



\section{Results}

\subsection{Academic test case}\label{subsec:academicTC}

\begin{figure}[h!]
\centering
\def\svgscale{0.5}
\begingroup%
  \makeatletter%
  \providecommand\color[2][]{%
    \errmessage{(Inkscape) Color is used for the text in Inkscape, but the package 'color.sty' is not loaded}%
    \renewcommand\color[2][]{}%
  }%
  \providecommand\transparent[1]{%
    \errmessage{(Inkscape) Transparency is used (non-zero) for the text in Inkscape, but the package 'transparent.sty' is not loaded}%
    \renewcommand\transparent[1]{}%
  }%
  \providecommand\rotatebox[2]{#2}%
  \newcommand*\fsize{\dimexpr\f@size pt\relax}%
  \newcommand*\lineheight[1]{\fontsize{\fsize}{#1\fsize}\selectfont}%
  \ifx\svgwidth\undefined%
    \setlength{\unitlength}{595.27559055bp}%
    \ifx\svgscale\undefined%
      \relax%
    \else%
      \setlength{\unitlength}{\unitlength * \real{\svgscale}}%
    \fi%
  \else%
    \setlength{\unitlength}{\svgwidth}%
  \fi%
  \global\let\svgwidth\undefined%
  \global\let\svgscale\undefined%
  \makeatother%
  \vspace{-6cm}
  \begin{picture}(1,1.41428571)%
    \lineheight{1}%
    \setlength\tabcolsep{0pt}%
    \put(0,0){\includegraphics[width=\unitlength,page=1]{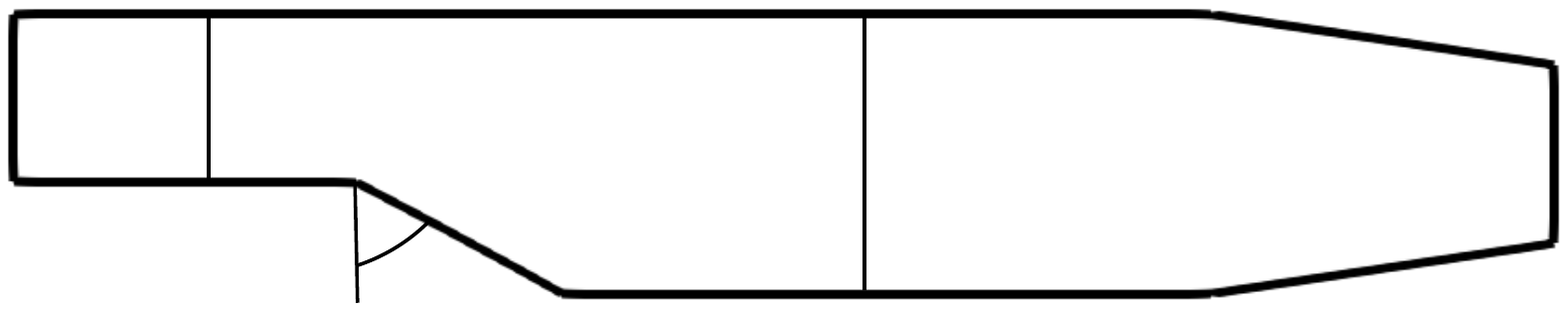}}%
    \put(0.11969246,0.772){\color[rgb]{0,0,0}\makebox(0,0)[lt]{\lineheight{0.5}\smash{\begin{tabular}[t]{l}$\Gamma_{in}$\end{tabular}}}}%
    \put(0.21,0.77){\color[rgb]{0,0,0}\makebox(0,0)[lt]{\begin{minipage}{0.08639455\unitlength}\raggedright $h_1$\end{minipage}}}%
    \put(0.29653131,0.68125705){\color[rgb]{0,0,0}\makebox(0,0)[lt]{\lineheight{0.5}\smash{\begin{tabular}[t]{l}$\mu$\end{tabular}}}}%
    \put(0.53681619,0.70060592){\color[rgb]{0,0,0}\makebox(0,0)[lt]{\lineheight{0.5}\smash{\begin{tabular}[t]{l}$h_2$\end{tabular}}}}%
    \put(0.79,0.73885345){\color[rgb]{0,0,0}\makebox(0,0)[lt]{\lineheight{0.5}\smash{\begin{tabular}[t]{l}$\Gamma_{out}$\end{tabular}}}}%
  \end{picture}%
  \vspace{-6cm}
\endgroup%

\caption{Geometry of the domain}\label{fig:shape}
\end{figure}

The first test case we propose to check the effectiveness of the procedure previously described is a classical 2D back step problem where the slope of the step is parametrized and can be varied (see \autoref{fig:shape}).

All the results provided in this paper are obtained by the use of an in-house open source library ITHACA-FV (In real Time Highly Advanced Computational Applications for Finite Volumes) \cite{RoSta17}, developed in a Finite Volume environment based on the solver OpenFOAM \cite{OpenFOAM}.

The set of equations we want to consider are the ones reported in \autoref{eq:rans} where $g_D = \left[1,0,0\right]^T$, $g_N = \bm{0}$, the eddy viscosity $\nu_t$ is obtained by the resolution of a $k-\epsilon$ turbulence model and $\nu=1\times10^{-3}$.

With reference to \autoref{fig:shape}, the height of the duct at the inlet, namely $h_1$, is equal to one while it is equal to $1.7$ in the middle of the channel, namely $h_2$. The domain is divided into $14\times10^3$ hexahedral cells mesh. The mesh motion is carried out by the use of a Radial Basis Function approach, as explained in \autoref{subsec:meshMotion}.

The Reynolds number characterizing the dynamics of the problem can be evaluated taking into account both the fluid properties together with geometrical aspects as:

\begin{equation*}
    Re = \frac{\overline{\bm{u}} \, h_2}{\nu} = 1.7 \times 10^3.
\end{equation*}

Since the range for the Reynolds number we are working at is on the border line between laminar and turbulent flows, we are forced to consider a turbulence closure model.

For the \textit{offline phase} we selected $50$ equispaced values of the parameter $\mu \in [0,75]$. Those values of the angle of the step were used to solve $50$ different full order problems in order to construct the snapshots matrix.

By applying a POD procedure, we can obtain the modal basis functions we need to project the equations. 

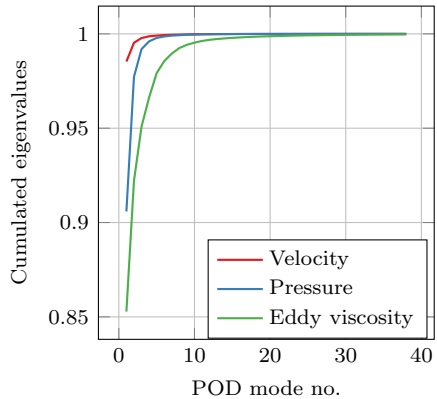
\begin{figure}[htbp]
    \centering
    \begin{tikzpicture}
        \begin{axis}[
            xlabel={POD mode no.},
            ylabel={Cumulated eigenvalues},
            legend style={at={(0.98,0.3)},anchor=north east},
            height=0.4\textwidth,
            width=0.4\textwidth, 
        ]
        \addplot+[no marks]
            table[x={xAx}, y={eigsU}] {backStepData/eigs.dat};
            
        \addplot+[no marks]
            table[x={xAx}, y={eigsP}] {backStepData/eigs.dat};
            
        \addplot+[no marks]
            table[x={xAx}, y={eigsNut}] {backStepData/eigs.dat};
        
        \legend{Velocity, Pressure, Eddy viscosity};
        \end{axis}
    \end{tikzpicture}
    \caption{Cumulated eigenvalues trends.}
    \label{fig:eigs}
\end{figure}

By analyzing \autoref{fig:eigs} we can notice that at least $25$ modes have to be selected for $\nu_t$ in order to catch the main part of the information contained into the offline snapshots. For what regards pressure and velocity manifolds, they are here projected and then reconstructed using $35$ basis functions.

Thus, a neural network has been constructed for the eddy viscosity approximation at every reduced SIMPLE algorithm step as explained in \autoref{subsec:rSimple}. 

The neural network employed here is composed by:
\begin{itemize}
    \item an input layer, whose dimension is equal to the dimension of the reduced velocity, i.e. $35$, plus one for the parameter;
    \item two hidden layers of dimension $256$ and $64$ respectively;
    \item an output layer of dimension $25$ for the reduced eddy viscosity coefficients.
\end{itemize}

The net is a fully connected one. Moreover the neurons of the hidden layers are characterized by the employment of ReLU activation functions. For the training procedure, the Adam optimizer has been selected and $10^4$ epochs have been fixed.

The training set is composed by both the intermediate and final solutions obtained during the offline phase, randomly selected. To control the training procedure, a test set has been selected too: $10$ totally random new parameter values have been chosen and their related full solutions have been calculated, saving both final and intermediate steps, coherently with the offline snapshots used for training.

\begin{figure}[htbp]
    \centering
    \begin{tikzpicture}
        \begin{semilogyaxis}[
            xlabel={Epochs},
            ylabel={Loss},
            legend style={at={(0.98,0.98)},anchor=north east},
            height=0.4\textwidth,
            width=0.4\textwidth, 
            x tick label style={
                               /pgf/number format/.cd,
                                    fixed,
                               /tikz/.cd
            },
            scaled x ticks=false
        ]
        \addplot+[no marks]
            table[x={xAxL}, y={trainLoss}] {backStepData/loss3525.dat};
            
        \addplot+[no marks]
            table[x={xAxL}, y={testLoss}] {backStepData/loss3525.dat};
        
        \legend{Train, Test};
        
        \end{semilogyaxis}
    \end{tikzpicture}
    \caption{Loss function decay for both train and test sets.}
    \label{fig:losses}
\end{figure}
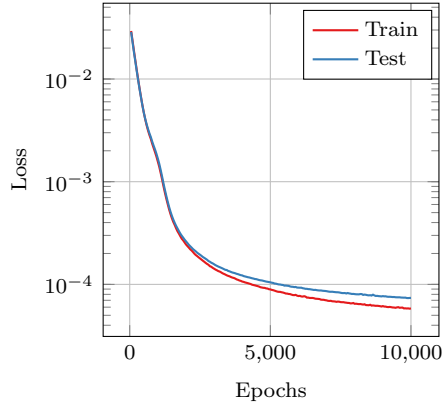

Looking at \autoref{fig:losses}, it can be noticed that there is a nice agreement between train and test loss functions. This is a good indicator for the extrapolation capability of the net.

\begin{figure}[htbp]
\centering
    \includegraphics[width=0.49 \textwidth]{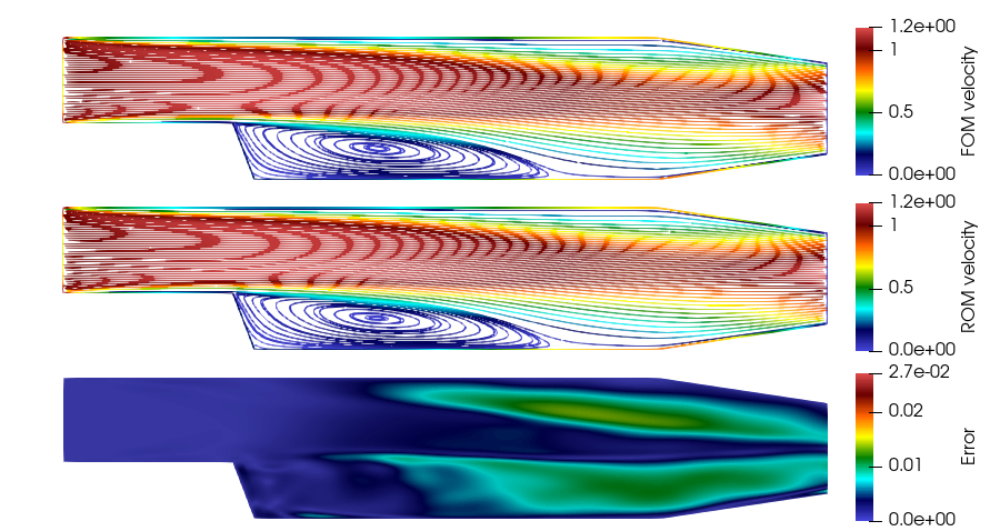}
    \includegraphics[width=0.49 \textwidth]{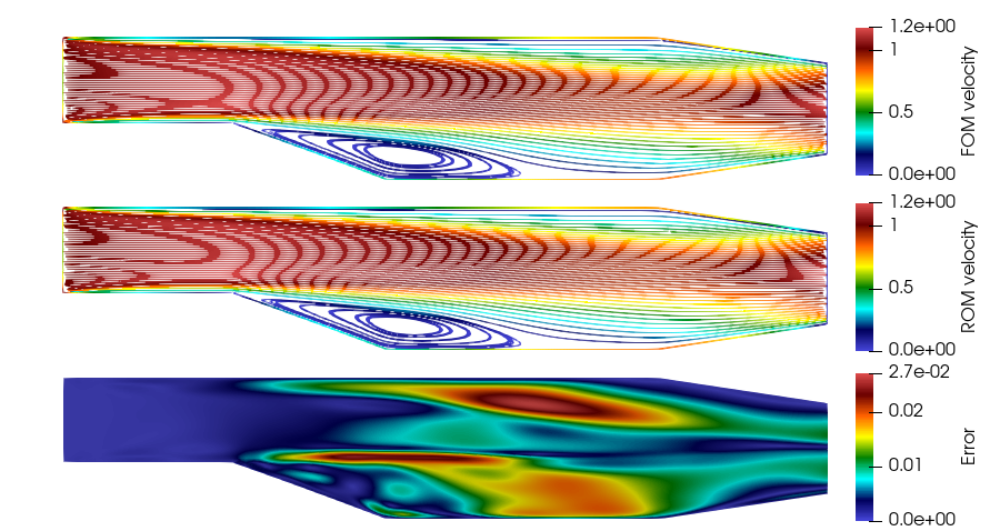}
    \caption{Comparison between velocity fields: FOM on top, ROM in the middle and error between them on bottom for $\mu=4.8$ and $\mu=68.3$.}
    \label{fig:velocities}
\end{figure}

\begin{figure}[htbp]
\centering
    \includegraphics[width=0.49 \textwidth]{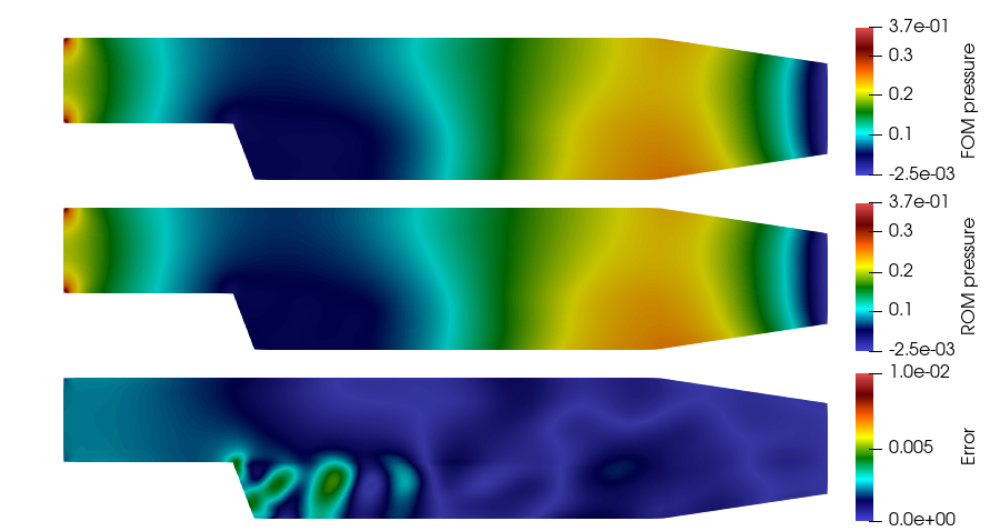}
    \includegraphics[width=0.49 \textwidth]{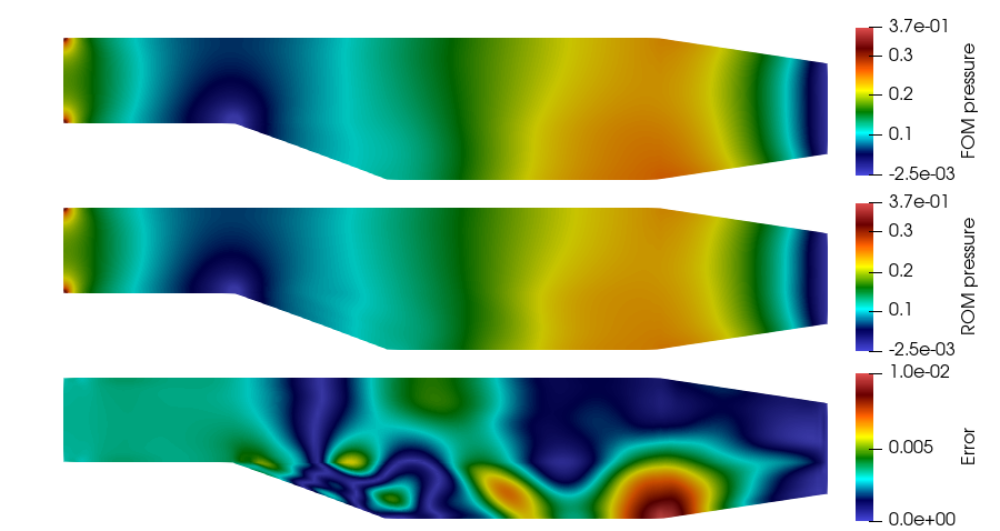}
    \caption{Comparison between pressure fields: FOM on top, ROM in the middle and error between them on bottom for $\mu=4.8$ and $\mu=68.3$.}
    \label{fig:pressures}
\end{figure}

\begin{figure}[htbp]
\centering
    \includegraphics[width=0.49 \textwidth]{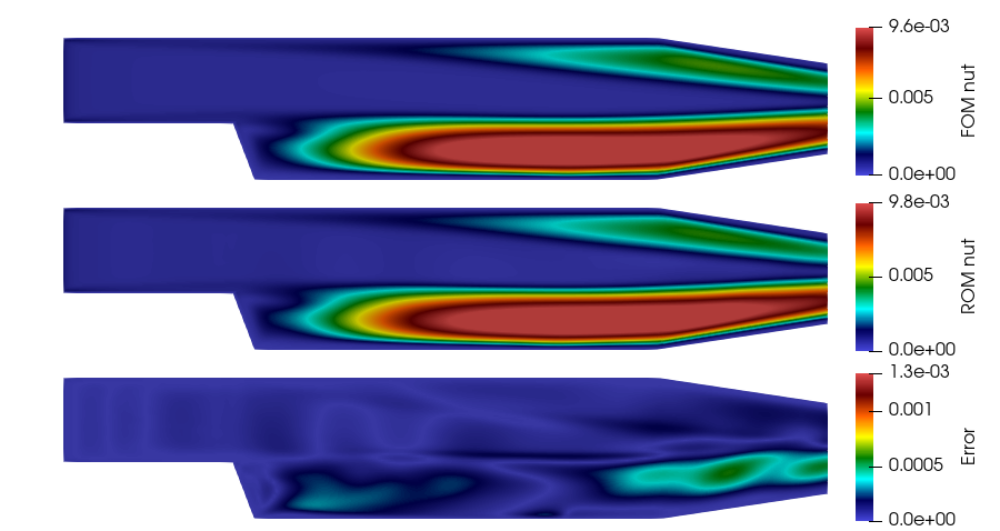}
    \includegraphics[width=0.49 \textwidth]{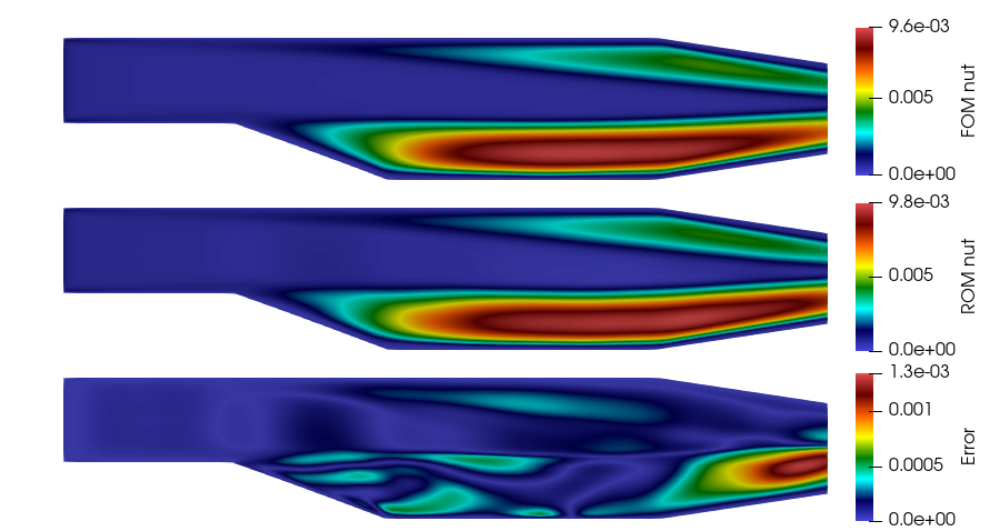}
    \caption{Comparison between eddy viscosity fields: FOM on top, ROM in the middle and error between them on bottom for $\mu=4.8$ and $\mu=68.3$.}
    \label{fig:nuts}
\end{figure}

In \autoref{fig:velocities}, \autoref{fig:pressures} and \autoref{fig:nuts}, we show the comparisons between full order model (FOM) and ROM solutions for velocity, pressure and eddy viscosity. Two random angles have been selected to show the behaviour of the model for both a very low parameter value and for a very high one.  

As it may be noticed, the reconstruction of the  reduced order model is very accurate and the errors are pretty low. The main differences between the high fidelity and the reduced solutions are present for high values of the parameter. This is to be addressed to the fact that the mesh is really distorted for those cases and the good orthogonality properties of the original mesh are lost. In any case the model is able to tackle the full order solution and can predict in a consistent way the correct solution.

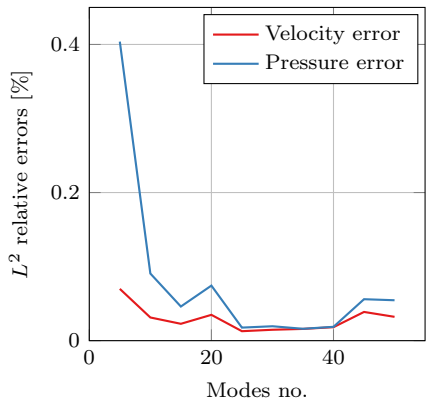
\begin{figure}[htbp]
    \centering
    \begin{tikzpicture}
        \begin{axis}[
            xlabel={Modes no.},
            ylabel={$L^2$ relative errors [\%]},
            ymin=0,
            ymax=0.45,
            xmin=0,
            xmax=55,
            legend style={at={(0.98,0.98)},anchor=north east},
            height=0.4\textwidth,
            width=0.4\textwidth,
        ]
        \addplot+[no marks]
            table[x={modes}, y={errorU}] {backStepData/errors.dat};
            
        \addplot+[no marks]
            table[x={modes}, y={errorP}] {backStepData/errors.dat};
        
        \legend{Velocity error, Pressure error};
        
        \end{axis}
    \end{tikzpicture}
    \caption{$L^2$ norm relative error for both velocity and pressure.}
    \label{fig:errors}
\end{figure}

As proof of what it has just been said, we show on \autoref{fig:errors} the trend of the $L^2$ norm relative errors while varying the dimension of the reduced manifolds for velocity and pressure at the same time. The values presented in this plot are the mean relative errors between $10$ random chosen parameters for the online phase.

\subsection{Ahmed body}\label{subsec:appliedTC}

\begin{figure}[htbp]
    \usetikzlibrary{backgrounds, calc}
     \def\gridstep{0.15}
    \centering
    \begin{subfigure}{0.45\textwidth}
         \centering
          \begin{tikzpicture}
             \node[] at (0, 0) {\includegraphics[trim={0 0 0 0},clip,width=1.\linewidth]{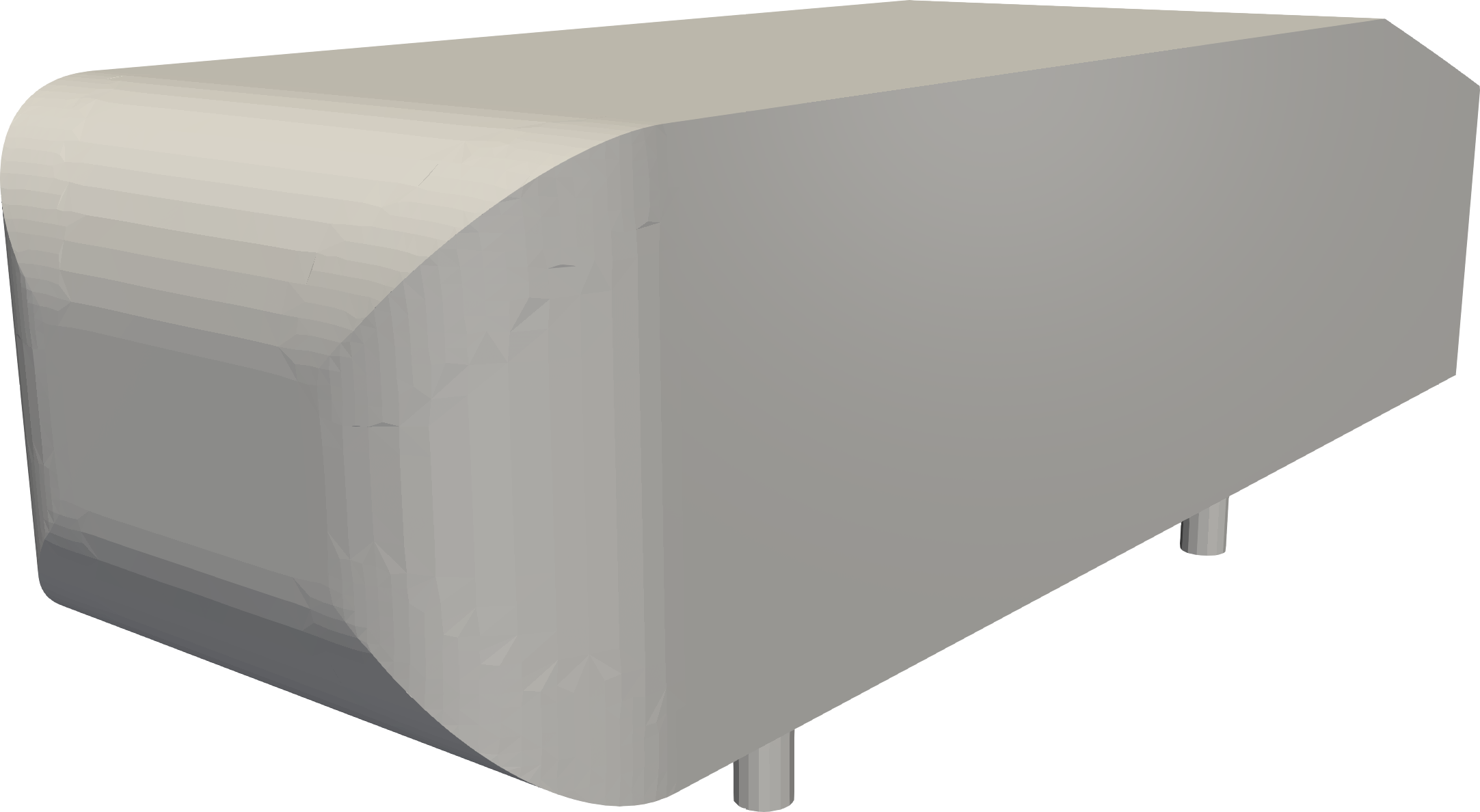}};
             \draw[very thin, overlay] ($(current bounding box.east) + (0.1, 3.8)$) -- ($(current bounding box.east) + (0.1, -3.8)$);
            \end{tikzpicture}
    \end{subfigure}\hfill%
    \begin{subfigure}{0.25\textwidth}
        \begin{tikzpicture}
             \node[] at (0, 0) {\includegraphics[trim={60cm 0 0 0},clip,width=1.\linewidth]{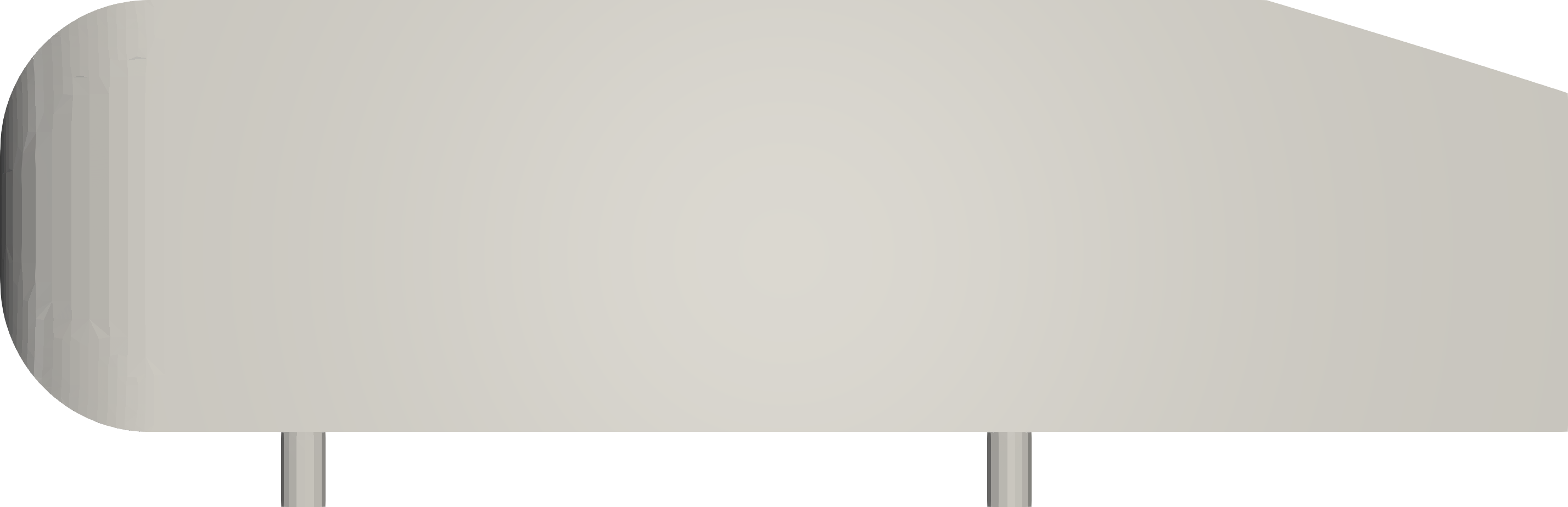}};
           \begin{pgfonlayer}{background}
                  \draw[help lines, thin, step=\gridstep, inner sep=0cm] ($(current bounding box.south west) + (0.05, 0.05)$) grid ($(current bounding box.north east) - (0.05, 0.05)$);
              \end{pgfonlayer}
        \end{tikzpicture} \\
        \begin{tikzpicture}
             \node[] at (0, 0) {\includegraphics[trim={60cm 0 0 0},clip,width=1.\linewidth]{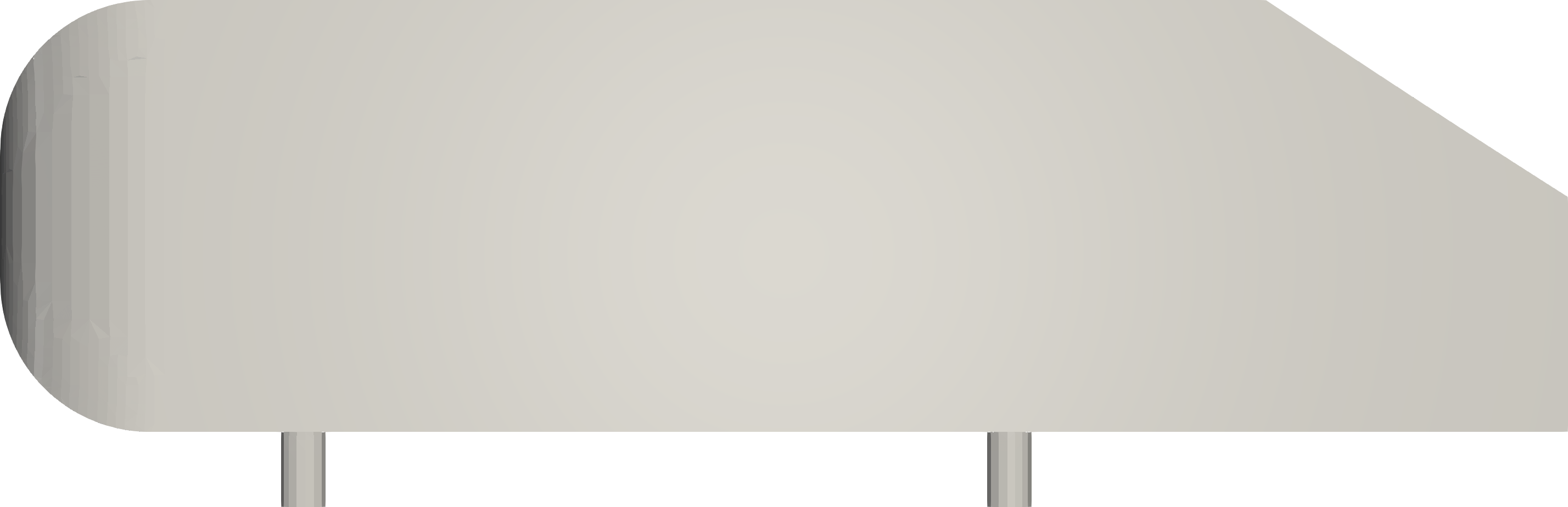}};
            \begin{pgfonlayer}{background}
                  \draw[help lines, thin, step=\gridstep, inner sep=0cm] ($(current bounding box.south west) + (0.05, 0.05)$) grid ($(current bounding box.north east) - (0.05, 0.05)$);
              \end{pgfonlayer}
        \end{tikzpicture} 
    \end{subfigure}
  \caption{Isometric view of the Ahmed body (left) and side views of the rear end with extreme values of the slant angle parameter (right). The minimum and maximum slant angles are \SI{15}{\degree} (top) and \SI{35}{\degree} (bottom), respectively.}
    \label{fig:ahmed_parameter}
\end{figure}
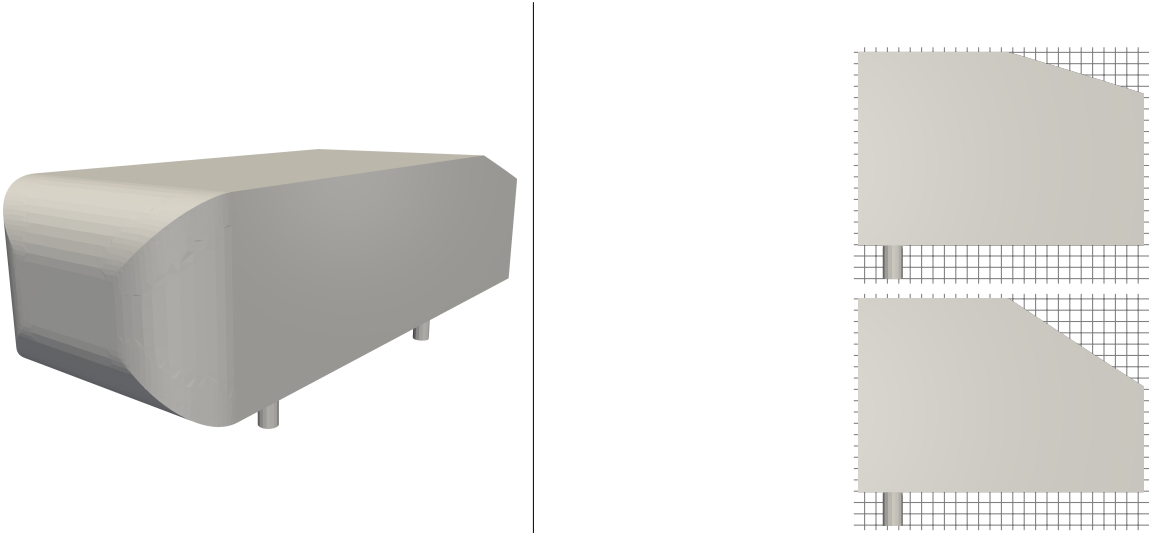

\begin{figure}[htbp]
    \centering
    \begin{tikzpicture}
        \begin{axis}[
             height=0.375\textwidth,
             width=0.75\textwidth,
            xlabel={Slant angle [\si{\degree}]},
            ylabel={$c_{d}$},
            legend style={at={(0.02,0.98)},anchor=north west},
            y tick label style={
                               /pgf/number format/.cd,
                                    fixed,
                                    fixed zerofill,
                                    precision=3,
                               /tikz/.cd
            },
        ]
        \addplot+[skip coords between index={10}{50}, only marks, mark=*, mark size=3pt]
            table[x={angles}, y={cd}] {ahmedData/tikz/raw_data/cds_offline.dat};
        \addplot+[skip coords between index={10}{50}, only marks, mark=o, mark size=3pt]
            table[x={angles}, y={cd}] {ahmedData/tikz/raw_data/cds_online.dat};
            \addplot+[skip coords between index={10}{50}, only marks, mark=x, mark size=3pt]
            table[x={angles}, y={cds_rom}] {ahmedData/tikz/raw_data/cds_online.dat};
        \addplot+[skip coords between index={0}{20}, mark=*, mark size=3pt, color=black!70]
            table[x={angles}, y={cds}] {ahmedData/tikz/raw_data/ahmed_RANS_cd.dat};
            \legend{FOM Train, FOM Test, ROM Test,};
        \end{axis}
    \end{tikzpicture}
    \caption{Drag coefficient $c_d$ over slant angle for the 20 full-order simulations: the even distribution of geometries into train and test sets is illustrated. For the test geometries, additionally, the ROM prediction is shown. In black, albeit not used in the present study, the development of the drag coefficients for higher slant angles is shown.}
    \label{fig:rom_and_fom_cd_over_angle}
\end{figure}
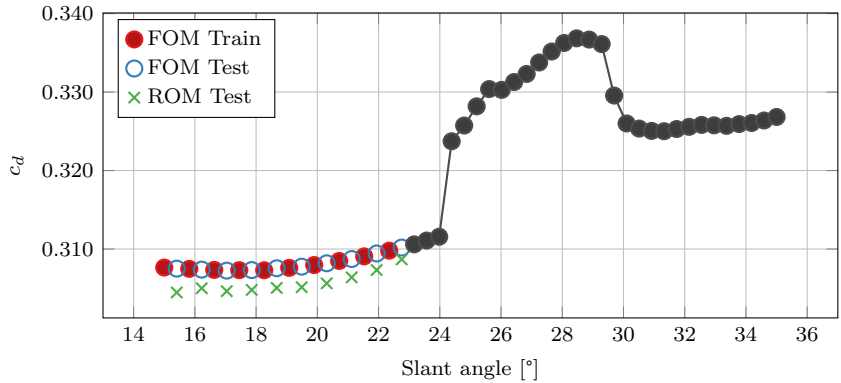

As the second test case, we chose an automotive external aerodynamic one: the Ahmed body \cite{ahmed1984}. The Ahmed body is a generic vehicle: the flow around the back of this bluff body contains the main flow structures that are encountered also for real-life vehicles. We defined one geometrical parameter – the slant angle  – using RBF mesh morphing (see Subsection \ref{subsec:meshMotion}). Figure \ref{fig:ahmed_parameter} shows the Ahmed body and illustrates the covered design space by the slant angle parameter. Depending on the slant angle, different flow regimes are encountered (cf.\ \autoref{fig:rom_and_fom_cd_over_angle}): (1) below approximately \SI{12}{\degree}, the flow remains attached over the slant; (2) between \SI{12}{\degree} and \SI{30}{\degree}, forming c-pillar vortices as well as recirculation regions at the top and base increase drag; (3) at approximately \SI{30}{\degree}, the flow fully separates off the slant, thus leading to a sudden drag decrease. At this stage, the study is restricted to the initial part of a single flow regime ranging from \SI{15}{\degree} to \SI{23}{\degree}, which already constitutes a demanding task.

We sampled the parameter range (\SI{15}{\degree} to \SI{23}{\degree}) uniformly with 20 RANS simulations using OpenFOAM\textsuperscript{\textregistered} with the Spalart-Allmaras turbulence model; these 20 simulations were decomposed into 10 for training (offline phase) the ROM and 10 to assess its accuracy (online phase). The inlet velocity for the simulations was set to \SI{40}{\meter\per\second}, thus resulting in a Reynolds number of $\approx$ $2.8 \times 10^6$  based on the model length. Each mesh was created with SnappyHexMesh\textsuperscript{\textregistered} and contained about \num{200000} cells; despite this small cell count, the typical flow regimes of the Ahmed body are encountered in the CFD solutions (cf.\  \autoref{fig:rom_and_fom_cd_over_angle}). While from a CFD perspective the meshes are very coarse, they constitute a challenge for the ROM and are considerably larger compared with those of the academic test case ($35 \times 10^4$ vs.\ $14 \times 10^3$). We saved every 20th of the total 2000 iterations as snapshots (velocity, pressure, and eddy viscosity fields), resulting in 100 snapshots per simulated slant angle. Each simulation took about 3 minutes on 16 CPU-cores. 

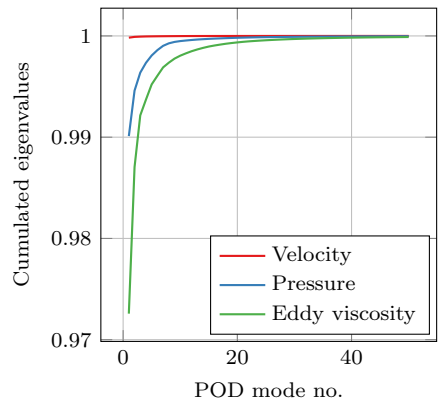
\begin{figure}[htbp]
    \centering
    \begin{tikzpicture}
    \begin{axis}[
        xlabel={POD mode no.},
        ylabel={Cumulated eigenvalues},
        legend style={at={(0.98,0.02)},anchor=south east},
        height=0.4\textwidth,
        width=0.4\textwidth, 
    ]
    \addplot+[no marks]
        table[x expr=\coordindex+1, y={ev_U}] {ahmedData/tikz/raw_data/cum_evs.dat};
        
   \addplot+[no marks]
        table[x expr=\coordindex+1, y={ev_p}] {ahmedData/tikz/raw_data/cum_evs.dat};
        
       \addplot+[no marks]
        table[x expr=\coordindex+1, y={ev_nut}] {ahmedData/tikz/raw_data/cum_evs.dat};
    
    \legend{Velocity, Pressure, Eddy viscosity};
    
    \end{axis}
    \end{tikzpicture}
    \caption{Cumulated eigenvalues of the POD for velocity, pressure, and eddy viscosity.}
    \label{fig:pod_cumulated_eigenvalues}
\end{figure}

After assembling the snapshot matrices with the intermediate as well as the converged iteration of the FOM simulations, we decomposed those matrices into modes and coefficients via POD. \autoref{fig:pod_cumulated_eigenvalues} shows the corresponding cumulated eigenvalues for velocity, pressure and eddy viscosity. For the upcoming investigations, we chose to keep 30 POD modes for all three fields. 

As described in \autoref{subsec:eddyViscosity}, the POD coefficients of the eddy viscosity are modeled via a neural network. For the present test case, the input of this neural network -- for each of the 1000 training samples (10 angle values times 100 saved iterations per angle) -- is given by the 30 POD coefficients of velocity and, additionally, the slant angle. The optimized neural network architecture consists of two hidden layers with 128 units each, Tanh activation functions, as well as a learning rate of $0.001$ for the Adam optimizer, thereby using a batch size of 128; the training was terminated after \num{10000} epochs. \\
Analogously as for the academic test case, we assessed the model accuracy on the test data set (the 1000 samples corresponding to the 10 test geometries) and found that the model generalizes well to unseen data.

With the trained neural network for the eddy viscosity, we are enabled to solve the reduced order problem for test geometries, i.e. slant angle configurations not present in the training data. Subsequently, we evaluate the ROM accuracy quantitatively and qualitatively by comparing ROM and FOM results for the 10 test geometries. For the quantitative analysis, we (1) compare the drag coefficients and (2) compare the relative $L^2$-errors between the velocity and pressure fields from ROM and FOM. For the qualitative comparison, we compare the velocity and pressure fields on two slices through the computational domain for two chosen test geometries.

We start the accuracy assessment with the drag coefficient, the major quantity of interest in the development of vehicle aerodynamics. As the drag coefficient of the ROM is obtained by integrating the pressure and wall shear stress over the vehicle surface, this investigation also allows to implicitly assess the accuracy of surface field predictions for those fields. \autoref{fig:rom_and_fom_cd_over_angle} shows the drag coefficient $c_d$ over the slant angle for the conducted 20 FOM simulations and indicates the even distribution in the parameter space of the geometries used for training and testing. 

The minimum and maximum absolute errors of the ROM are 1.5 (test sample at slant angle \SI{22.8}{\degree}) and 3.0 (\SI{15.4}{\degree}) drag counts, respectively, while the mean error over all 10 test samples amounts to 2.4 drag counts. The drag coefficient in automotive vehicle aerodynamics is dominated by the pressure contribution (approximately \SI{85}{\percent} pressure and \SI{15}{\percent} viscous contribution for the present test case); accordingly, we found that the error in surface pressure between ROM and FOM accounts for the majority of the total error in the drag coefficient prediction. Therefore, the visible systematic offset between ROM and FOM for the drag coefficient can probably be reduced by improving the pressure field prediction, which is investigated next.

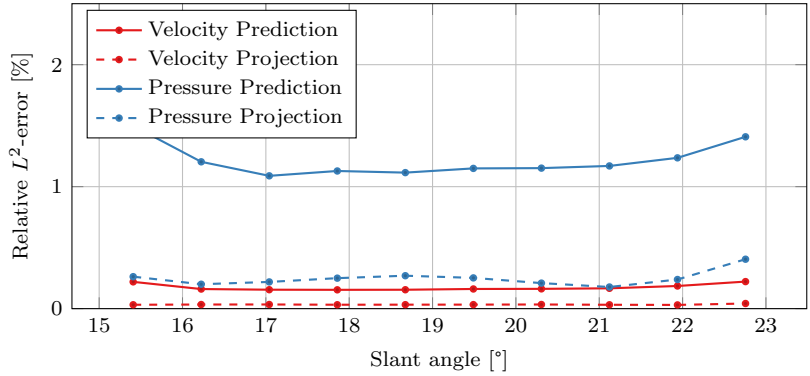
\begin{figure}[htbp]
    \centering
    \begin{tikzpicture}
        \begin{axis}[
            xlabel={Slant angle [\si{\degree}]},
            ylabel={Relative $L^2$-error [\%]},
            ymin=0., 
            ymax=2.5,
            legend style={at={(0.02,0.98)},anchor=north west},
        ]
        \addplot+[skip coords between index={10}{50}, mark=*, mark size=1pt]
            table[x={angles}, y={errU}] {ahmedData/tikz/raw_data/errors_fields.dat};
        \pgfplotsset{cycle list shift=-1}
        \addplot+[skip coords between index={10}{50}, mark=*, dashed, mark size=1pt]
            table[x={angles}, y={projerrsU}] {ahmedData/tikz/raw_data/errors_fields.dat};
        \addplot+[skip coords between index={10}{50}, mark=*, mark size=1pt]
            table[x={angles}, y={errP}] {ahmedData/tikz/raw_data/errors_fields.dat};
        \pgfplotsset{cycle list shift=-2}
         \addplot+[skip coords between index={10}{50}, mark=*, dashed, mark size=1pt]
            table[x={angles}, y={projerrsP}] {ahmedData/tikz/raw_data/errors_fields.dat};
          \legend{Velocity Prediction, Velocity Projection, Pressure Prediction, Pressure Projection};
        \end{axis}
    \end{tikzpicture}
    \caption{Quantitative errors of the ROM predictions for velocity and pressure fields of the test samples (cf. Figure \ref{fig:rom_and_fom_cd_over_angle}). The ROM errors (solid lines) lines are compared with those from the projection of the FOM solution into the POD subspace (dashed lines).}
    \label{fig:quantitative_errors_fields}
\end{figure}

\autoref{fig:quantitative_errors_fields} shows the relative $L^2$-errors between ROM prediction and FOM (solid lines) for velocity and pressure. As for the drag coefficient, the highest errors for both fields are encountered for the test sample with \SI{15.4}{\degree} slant angle. The errors for pressure are one magnitude higher compared with those for velocity. Additionally, the projection errors -- the lower bounds for the ROM errors --  are shown (dashed lines). While for the velocity a ROM prediction error close to the projection error is achieved, there is still room for improvement in the case of pressure (vertical distance between blue solid and dashed lines).

\begin{figure}[htbp]
    \centering
    \begin{subfigure}{0.36\textwidth}
    \captionsetup{position=top}
    \centering
     \includegraphics[trim={0 0 0 0},clip,height=1.97cm]{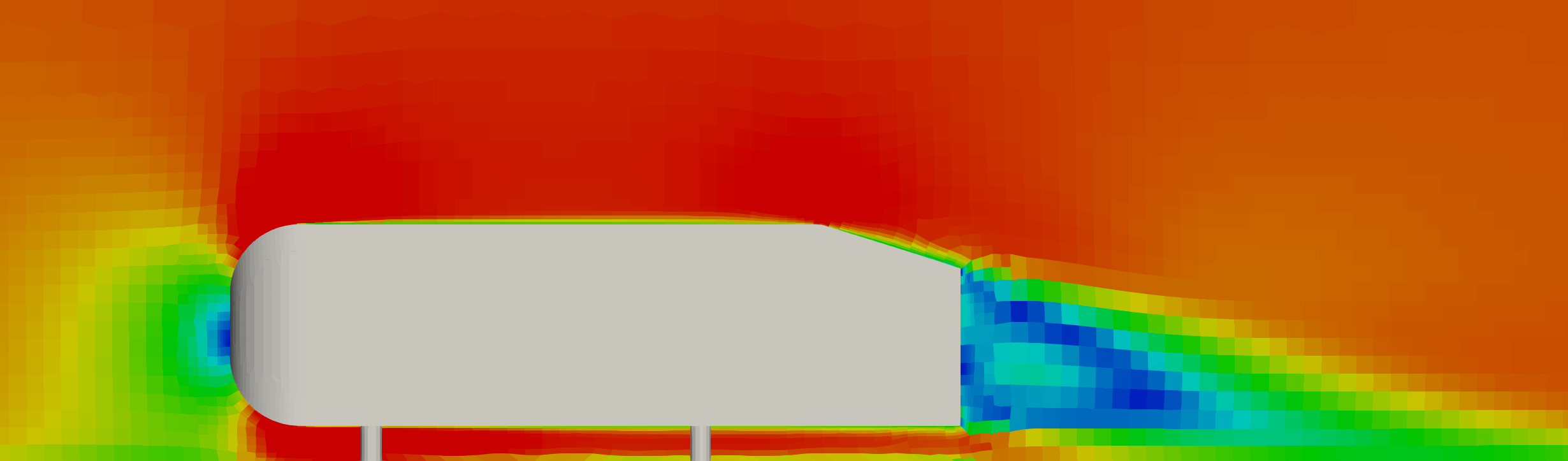}
    \end{subfigure}\hfill%
    \begin{subfigure}{0.36\textwidth}
    \centering
    \includegraphics[trim={0 0 0 0}, clip,height=1.97cm]{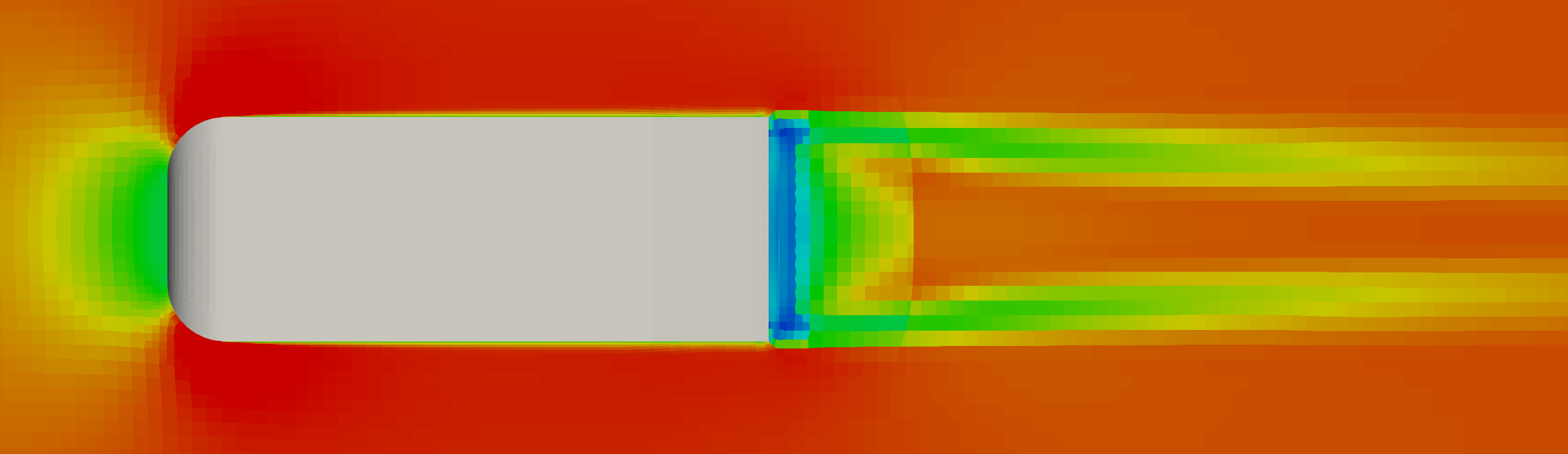}
    \end{subfigure}%

    \begin{subfigure}{0.36\textwidth}
    \captionsetup{position=top}
    \centering
     \includegraphics[trim={0 0 0 0},clip,height=1.97cm]{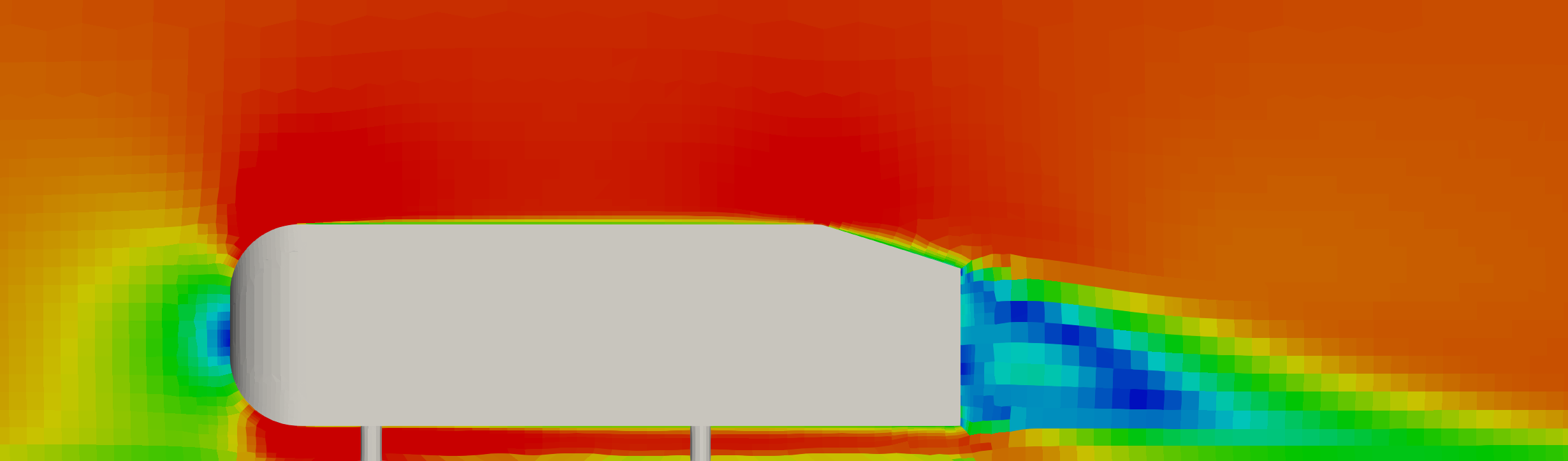}
    \end{subfigure}\hfill%
    \begin{subfigure}{0.36\textwidth}
    \centering
   \includegraphics[trim={0 0 0 0}, clip,height=1.97cm]{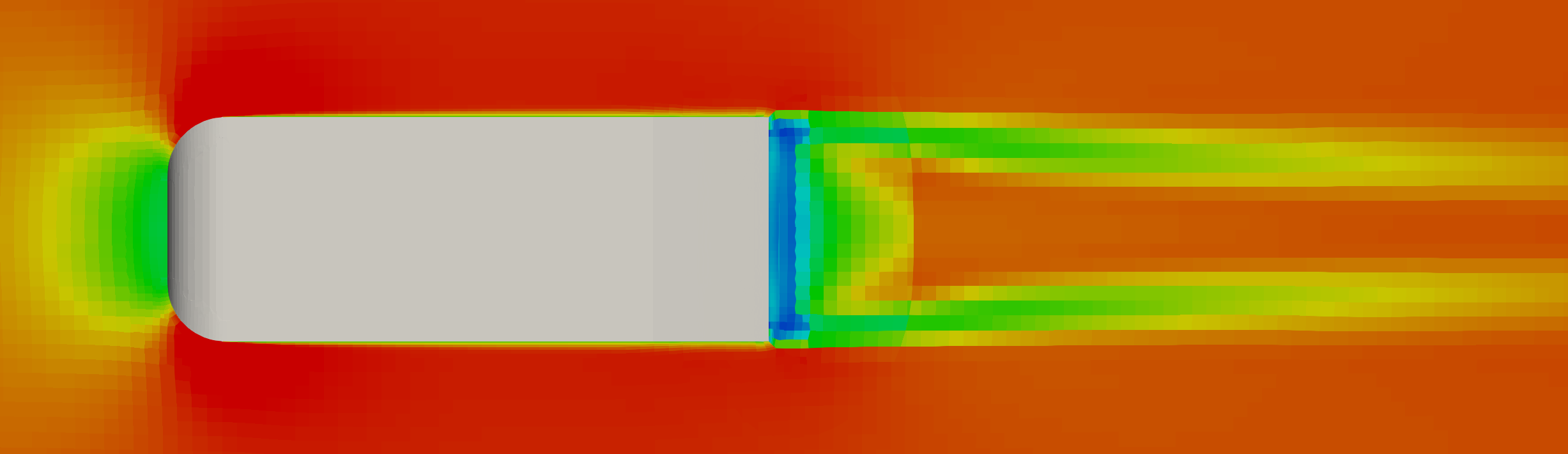}
    \end{subfigure}
    
    \begin{subfigure}{0.36\textwidth}
    \captionsetup{position=top}
    \centering
     \includegraphics[trim={0 0 0 0},clip,height=1.97cm]{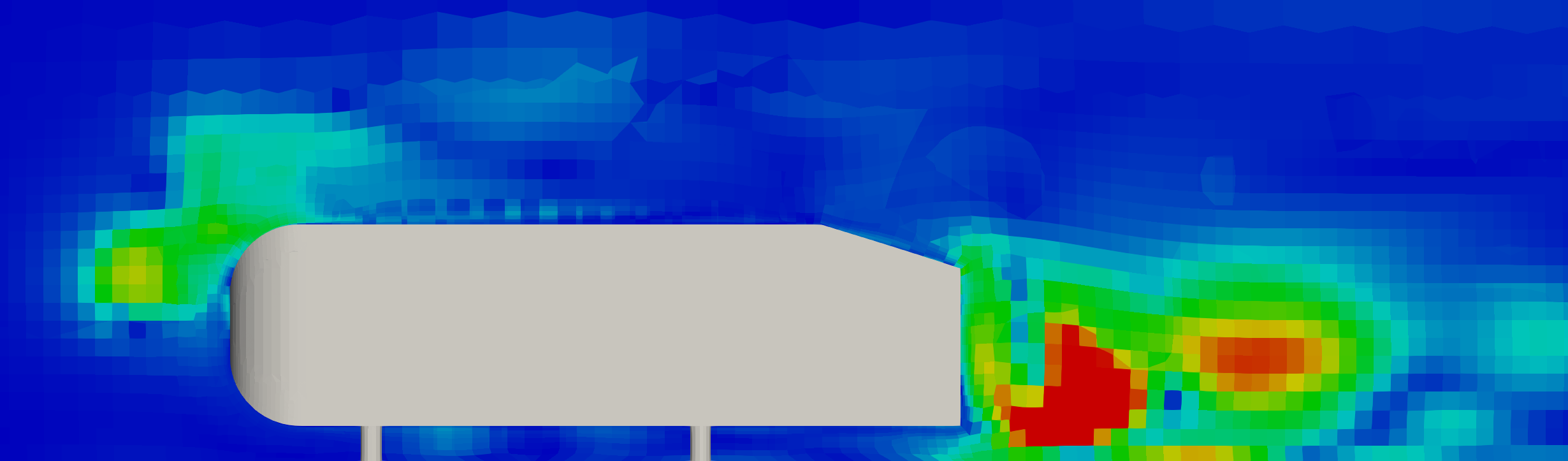}
    \end{subfigure}\hfill%
    \begin{subfigure}{0.36\textwidth}
    \centering
    \includegraphics[trim={0 0 0 0}, clip,height=1.97cm]{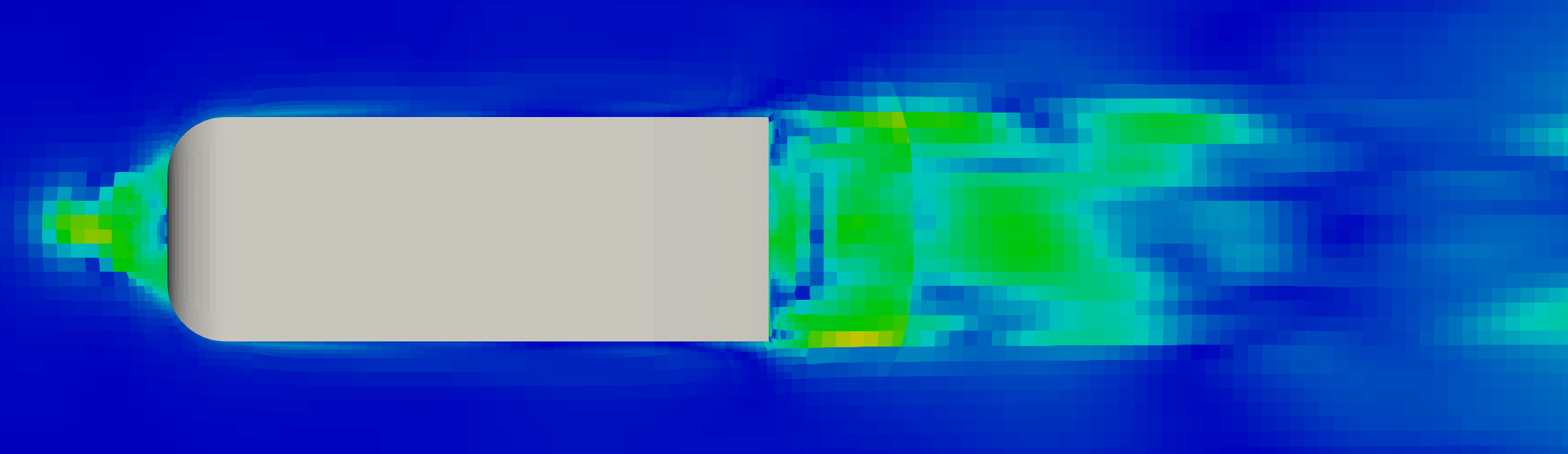}
    \end{subfigure}%

     \begin{subfigure}{0.99\textwidth}
    \centering
    \caption{Test sample with slant angle \SI{15.4}{\degree}}
    \end{subfigure}
    \vspace{0.7em}

    \begin{subfigure}{0.36\textwidth}
    \centering
         \includegraphics[trim={0 0 0 0},clip,height=1.97cm]{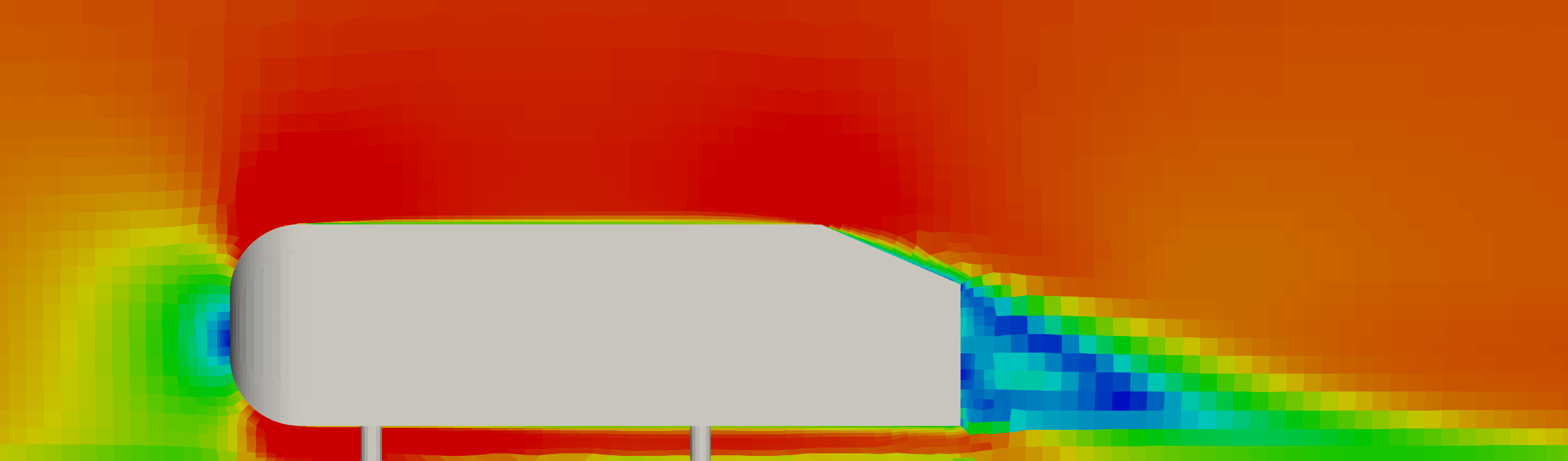}
    \end{subfigure}\hfill%
    \begin{subfigure}{0.36\textwidth}
    \centering
      \includegraphics[trim={0 0 0 0},clip,height=1.97cm]{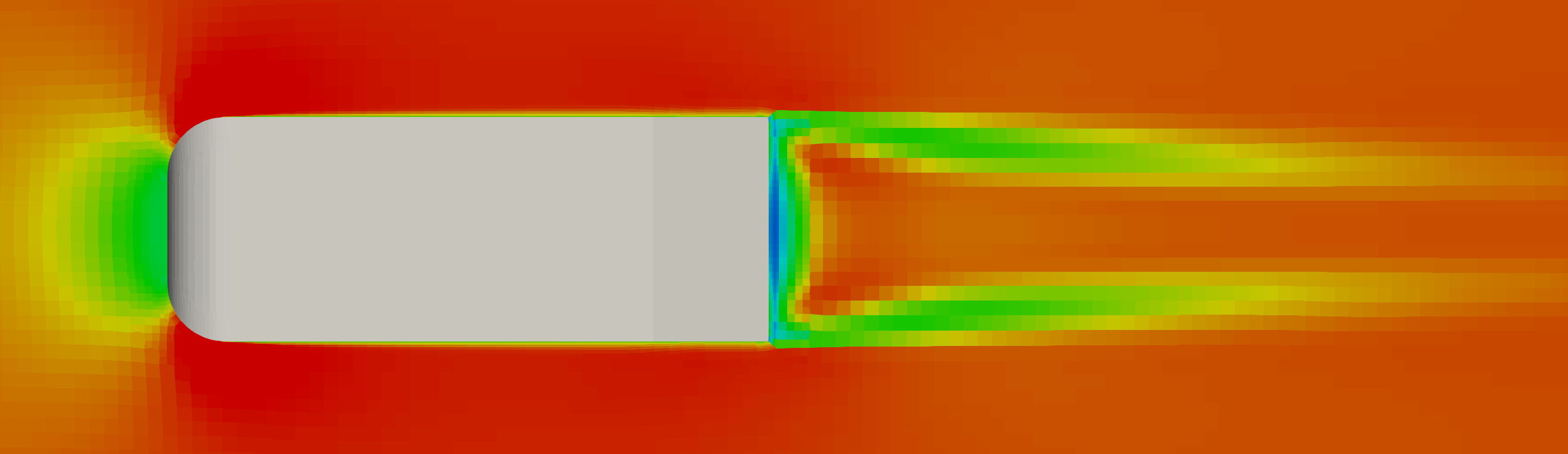}
    \end{subfigure}%
    
    \begin{subfigure}{0.36\textwidth}
    \centering
         \includegraphics[trim={0 0 0 0},clip,height=1.97cm]{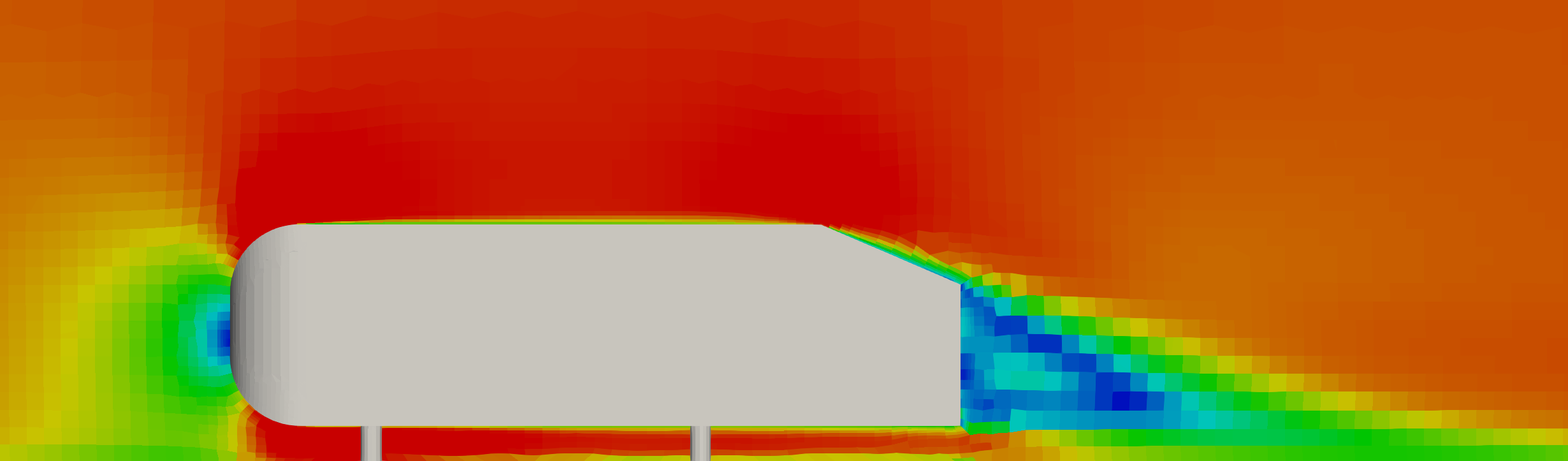}
    \end{subfigure}\hfill%
    \begin{subfigure}{0.36\textwidth}
    \centering
      \includegraphics[trim={0 0 0 0},clip,height=1.97cm]{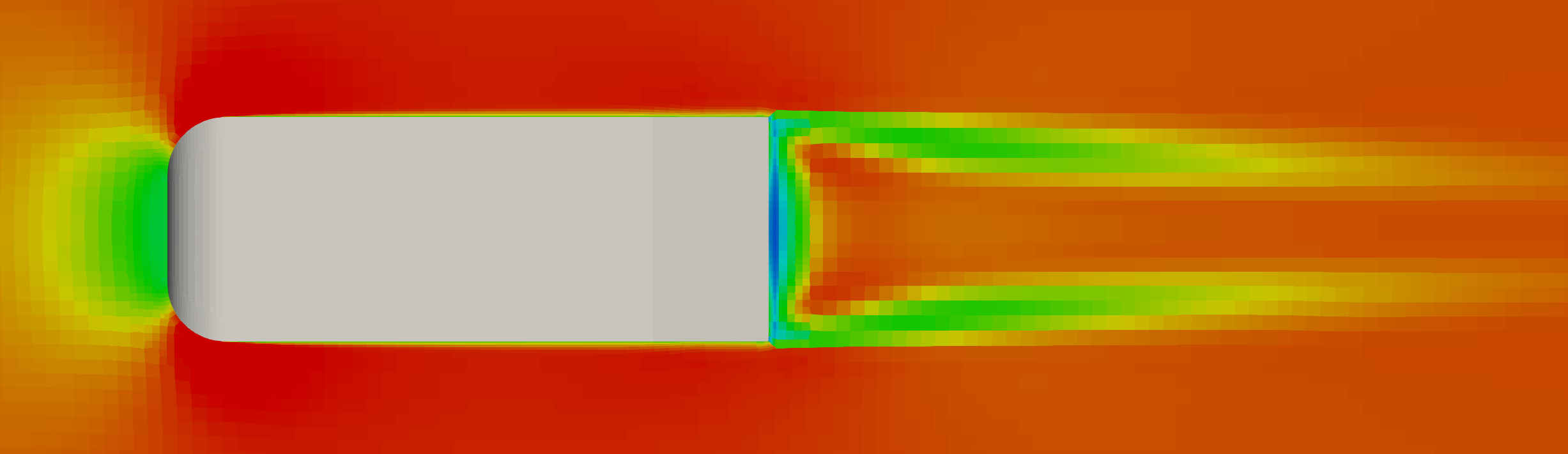}
    \end{subfigure}%
    
    \begin{subfigure}{0.36\textwidth}
    \centering
         \includegraphics[trim={0 0 0 0},clip,height=1.97cm]{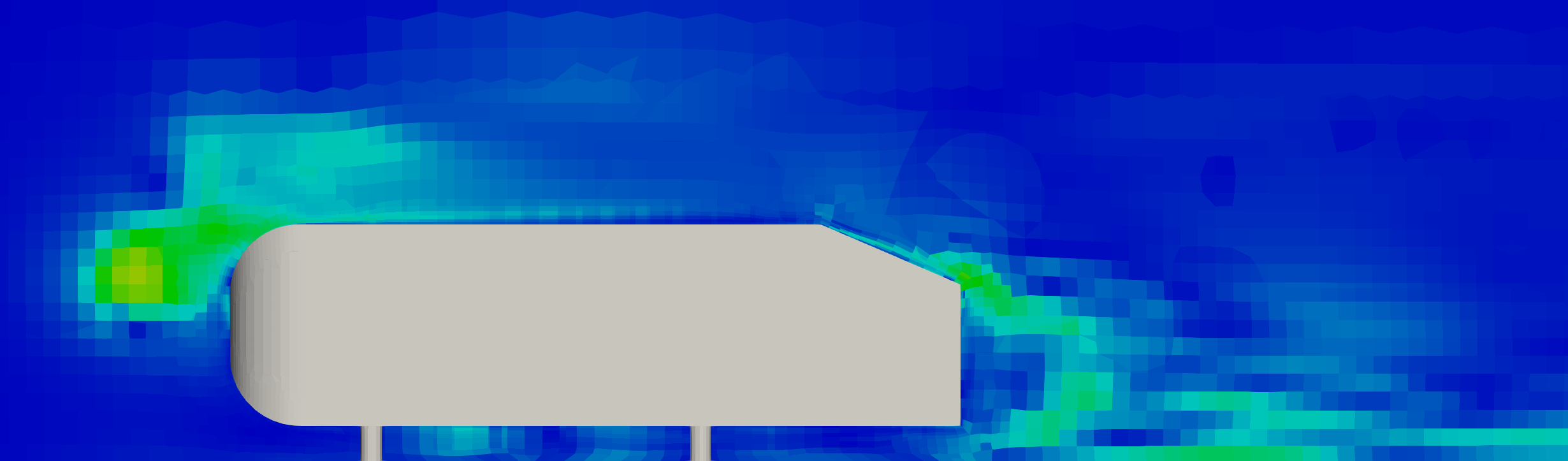}
    \end{subfigure}\hfill%
    \begin{subfigure}{0.36\textwidth}
    \centering
       \includegraphics[trim={0 0 0 0},clip,height=1.97cm]{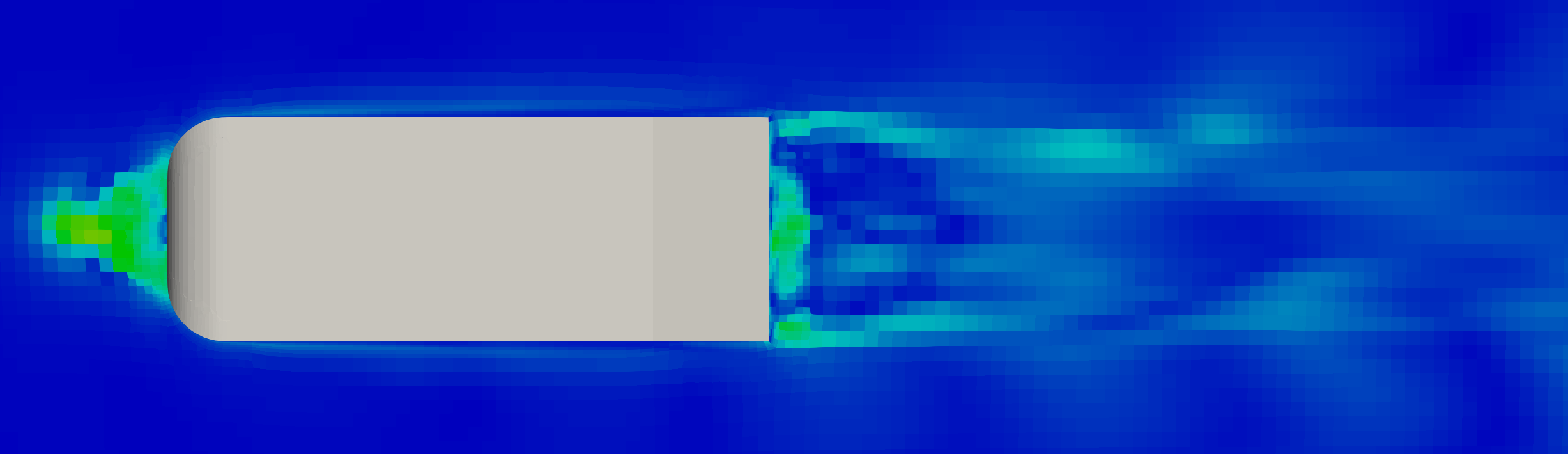}
    \end{subfigure}%

     \begin{subfigure}{0.99\textwidth}
    \centering
    \caption{Test sample with slant angle \SI{22.8}{\degree}}
    \end{subfigure}
      \vspace{0.7em}
    
     \begin{subfigure}{0.36\textwidth}
    \centering
     \begin{tikzpicture}
    \pgfplotscolorbardrawstandalone[ 
        colormap/bluered,
        colorbar sampled,
        colorbar horizontal,
        point meta min=0,
        point meta max=45,
        colorbar style={
             font=\footnotesize,
            title=Velocity Magnitude[\si{\meter \per \second}],
            width=6cm,
            samples=256,}]
    \end{tikzpicture}
    \end{subfigure}
     \begin{subfigure}{0.36\textwidth}
    \centering
     \begin{tikzpicture}
    \pgfplotscolorbardrawstandalone[ 
        colormap/bluered,
        colorbar sampled,
        colorbar horizontal,
        point meta min=0,
        point meta max=1,
        colorbar style={
             font=\footnotesize,
            title=Velocity Difference Magnitude [\si{\meter \per \second}],
            width=6cm,
            samples=256,}]
    \end{tikzpicture}
    \end{subfigure}
    
    \caption{Qualitative comparison for the velocity on the centerplane (left) and a slice $\SI{0.24}{\meter}$ above the street (right): FOM results (top), ROM predictions (middle), and difference ($\text{ROM} - \text{FOM}$, bottom) for the test sample with lowest (a) and highest (b) slant angle.}
     \label{fig:rom_U_qualitative_comparison}
\end{figure}

\begin{figure}[htbp]
    \centering
    \begin{subfigure}{0.36\textwidth}
    \captionsetup{position=top}
    \centering
     \includegraphics[trim={0 0 0 0},clip,height=1.97cm]{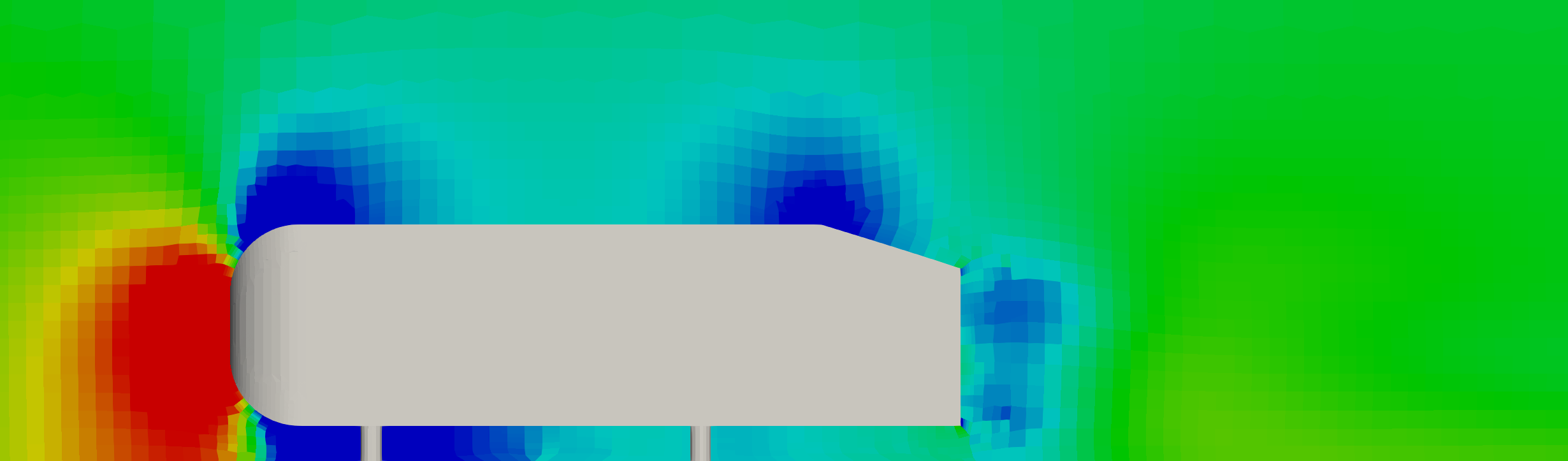}
    \end{subfigure}\hfill%
    \begin{subfigure}{0.36\textwidth}
    \centering
    \includegraphics[trim={0 0 0 0}, clip,height=1.97cm]{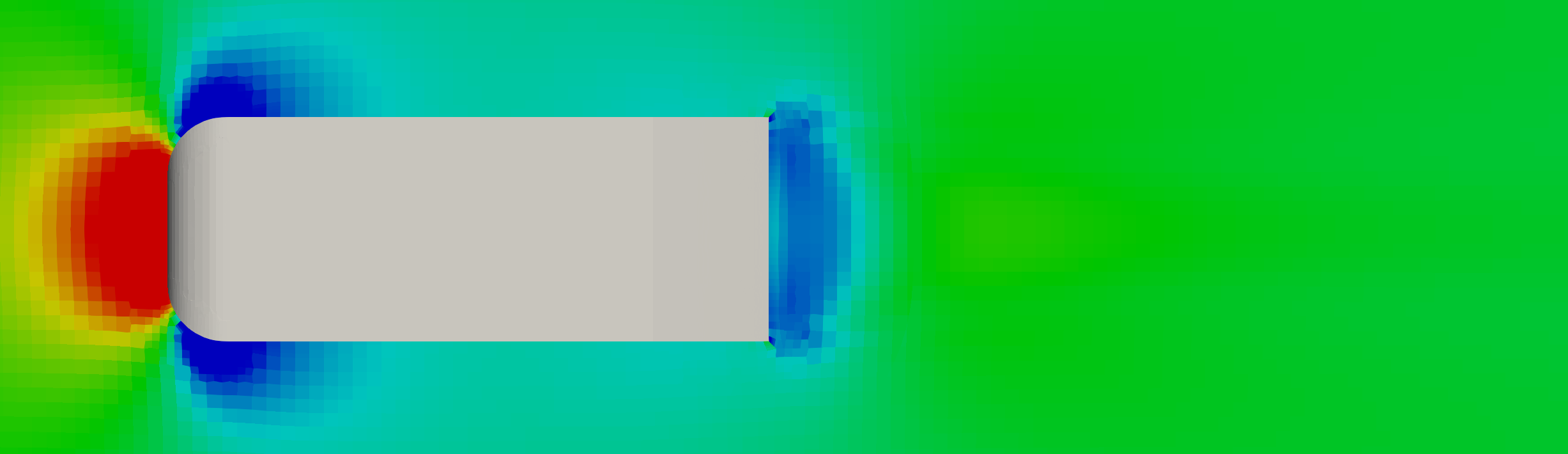}
    \end{subfigure}%

    \begin{subfigure}{0.36\textwidth}
    \captionsetup{position=top}
    \centering
     \includegraphics[trim={0 0 0 0},clip,height=1.97cm]{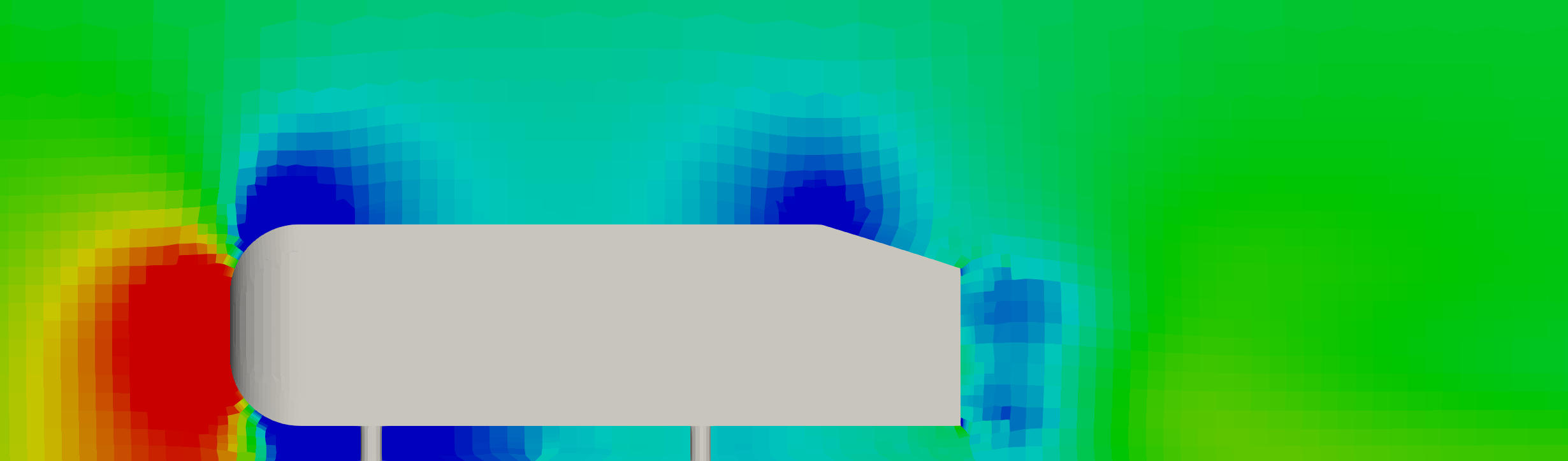}
    \end{subfigure}\hfill%
    \begin{subfigure}{0.36\textwidth}
    \centering
   \includegraphics[trim={0 0 0 0}, clip,height=1.97cm]{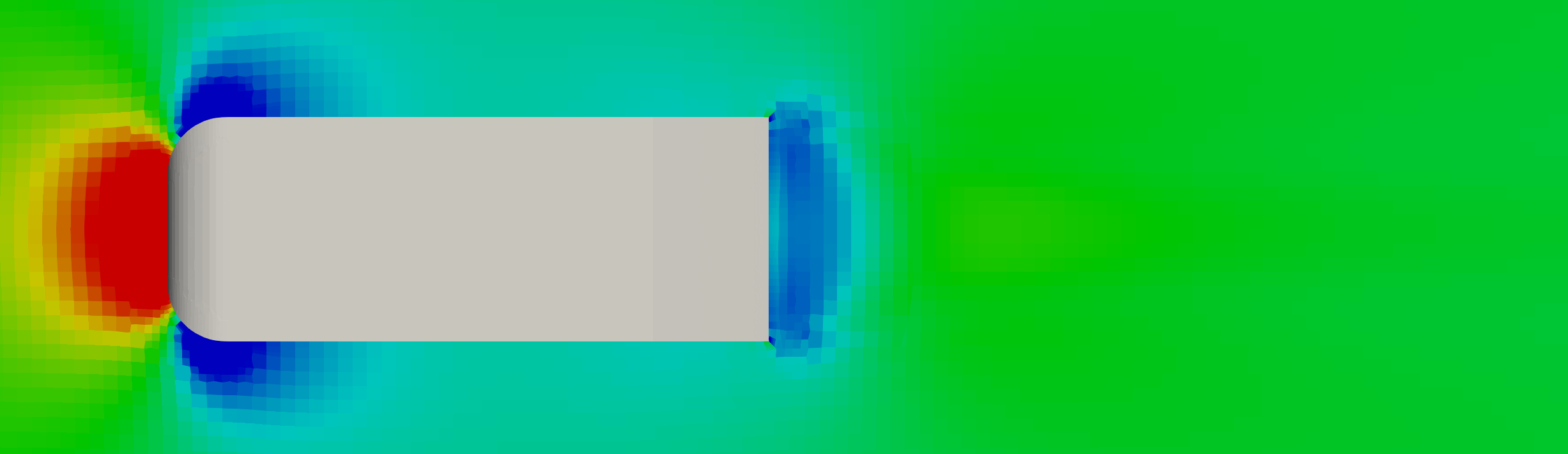}
    \end{subfigure}
    
    \begin{subfigure}{0.36\textwidth}
    \captionsetup{position=top}
    \centering
     \includegraphics[trim={0 0 0 0},clip,height=1.97cm]{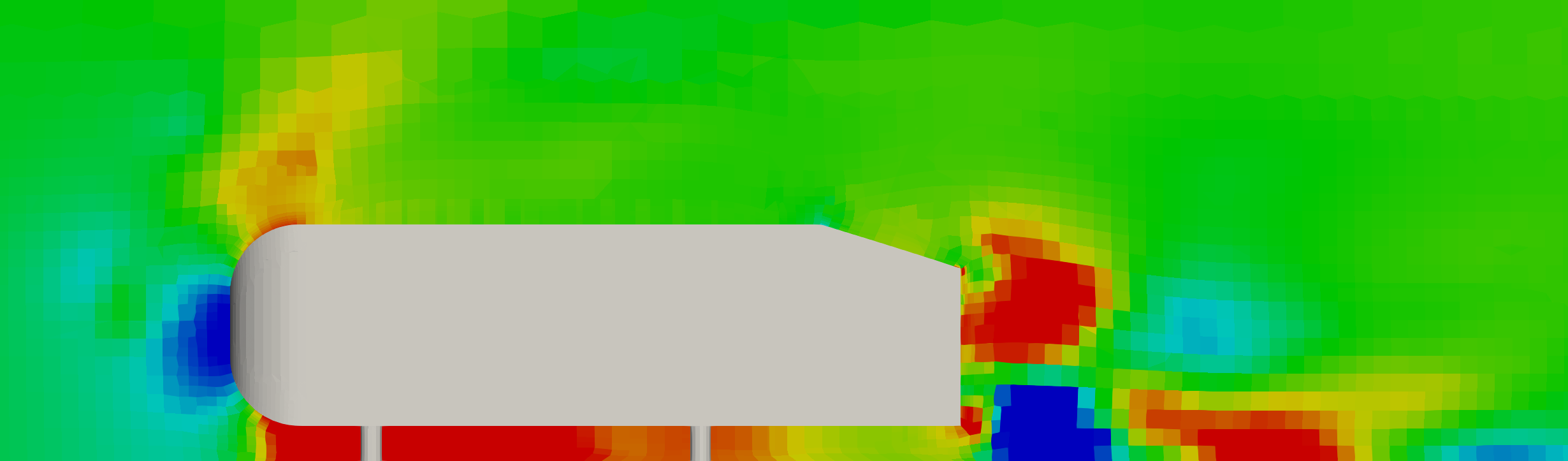}
    \end{subfigure}\hfill%
    \begin{subfigure}{0.36\textwidth}
    \centering
    \includegraphics[trim={0 0 0 0}, clip,height=1.97cm]{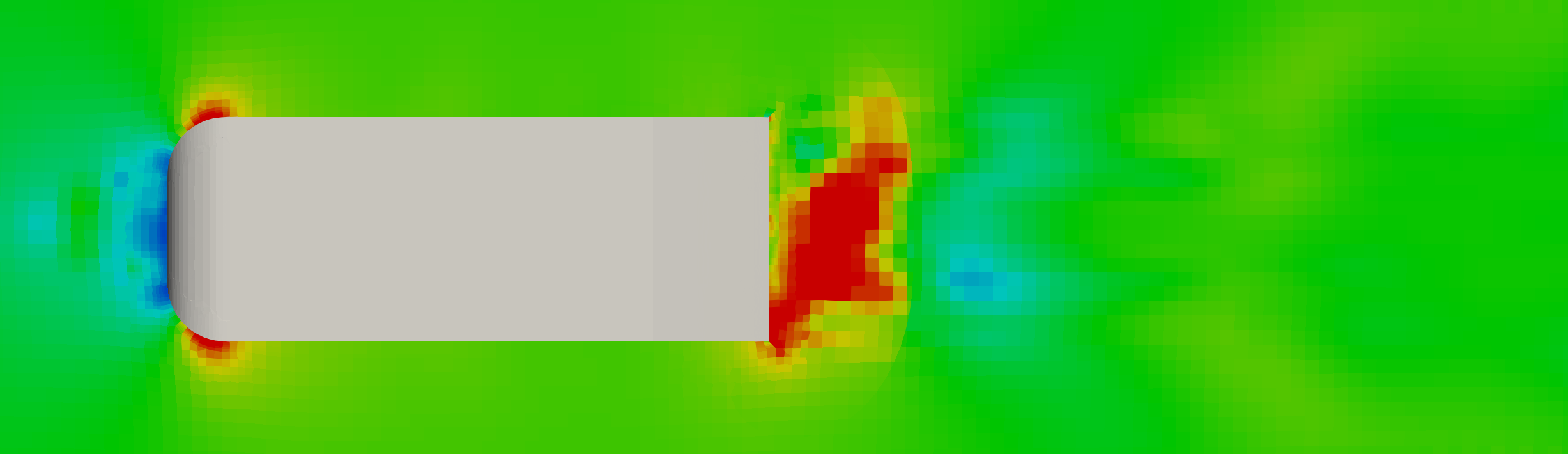}
    \end{subfigure}%

     \begin{subfigure}{0.99\textwidth}
    \centering
    \caption{Test sample with slant angle \SI{15.4}{\degree}}
    \end{subfigure}
    \vspace{0.7em}

    \begin{subfigure}{0.36\textwidth}
    \centering
         \includegraphics[trim={0 0 0 0},clip,height=1.97cm]{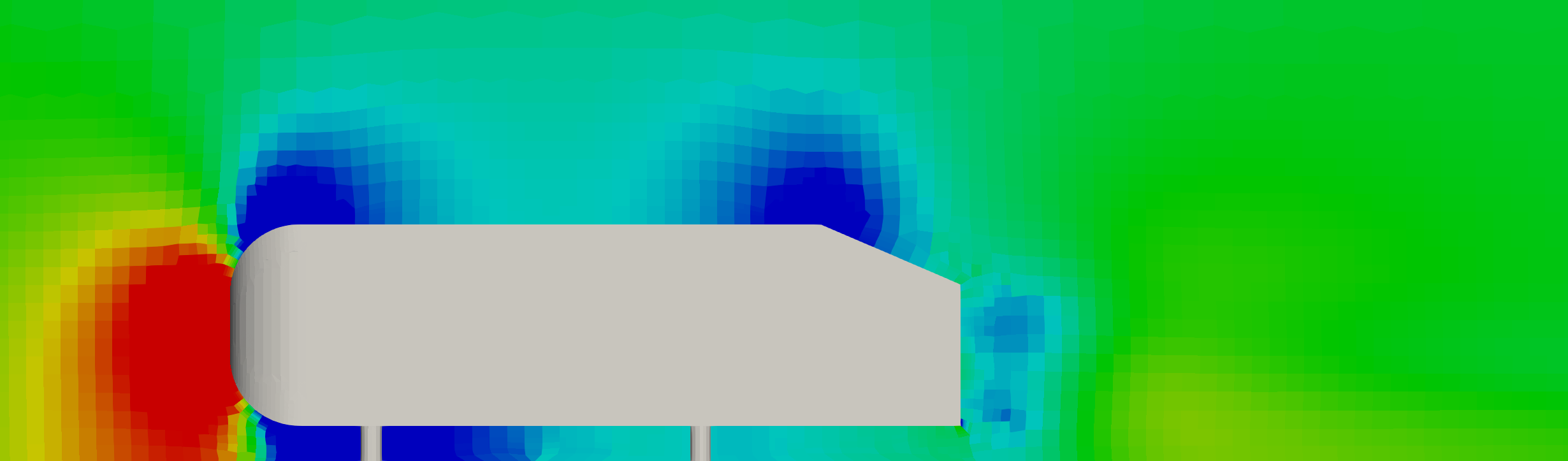}
    \end{subfigure}\hfill%
    \begin{subfigure}{0.36\textwidth}
    \centering
      \includegraphics[trim={0 0 0 0},clip,height=1.97cm]{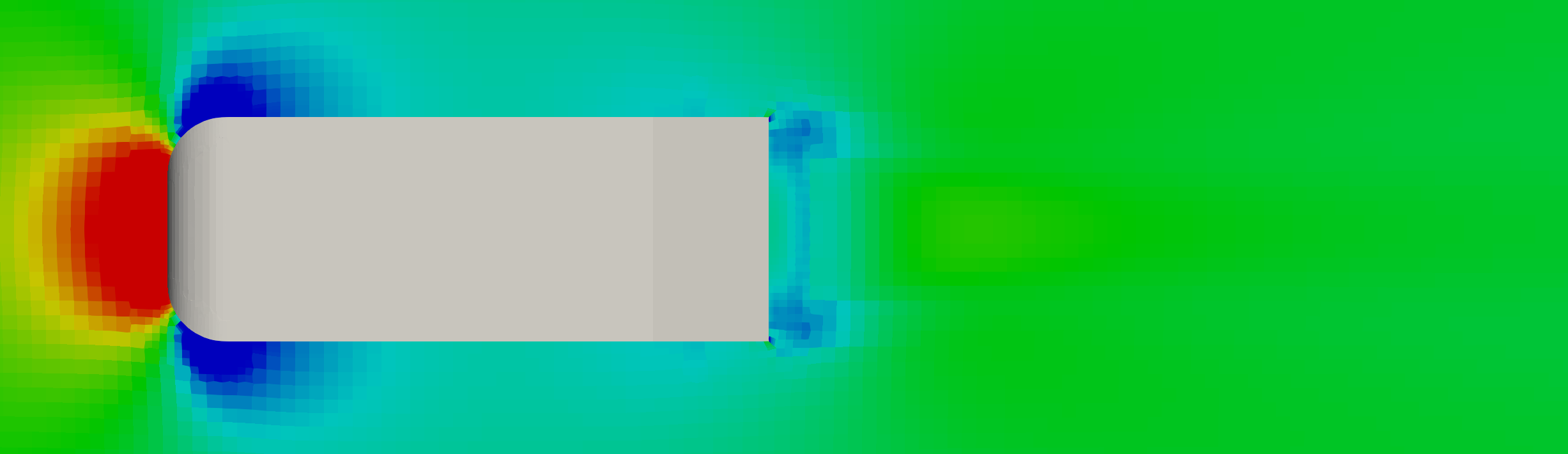}
    \end{subfigure}%
    
    \begin{subfigure}{0.36\textwidth}
    \centering
         \includegraphics[trim={0 0 0 0},clip,height=1.97cm]{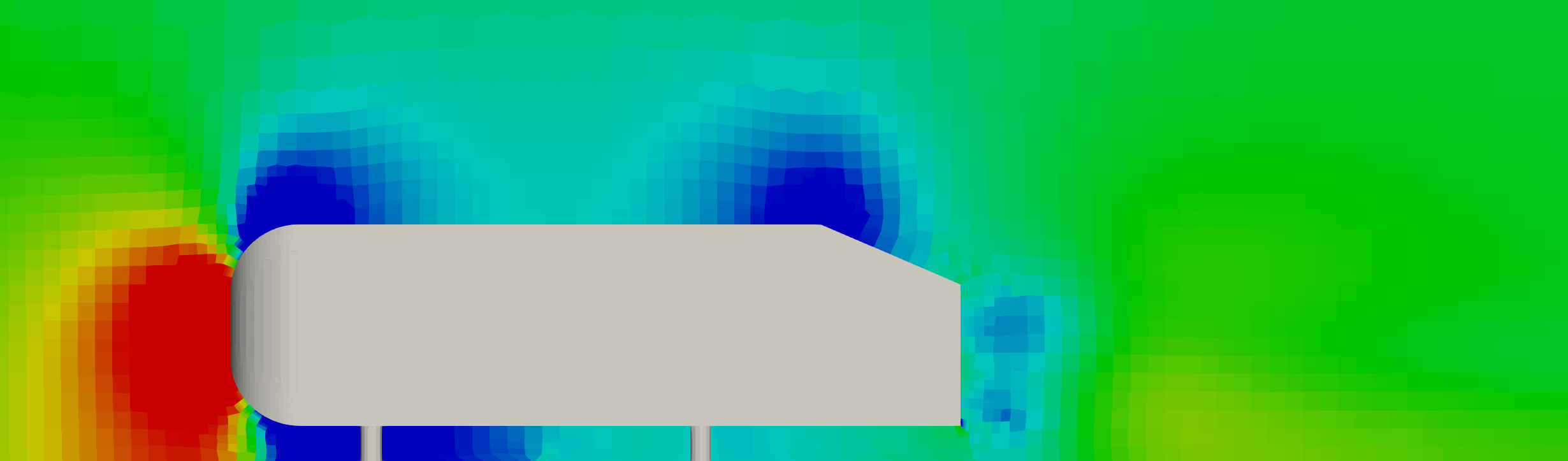}
    \end{subfigure}\hfill%
    \begin{subfigure}{0.36\textwidth}
    \centering
      \includegraphics[trim={0 0 0 0},clip,height=1.97cm]{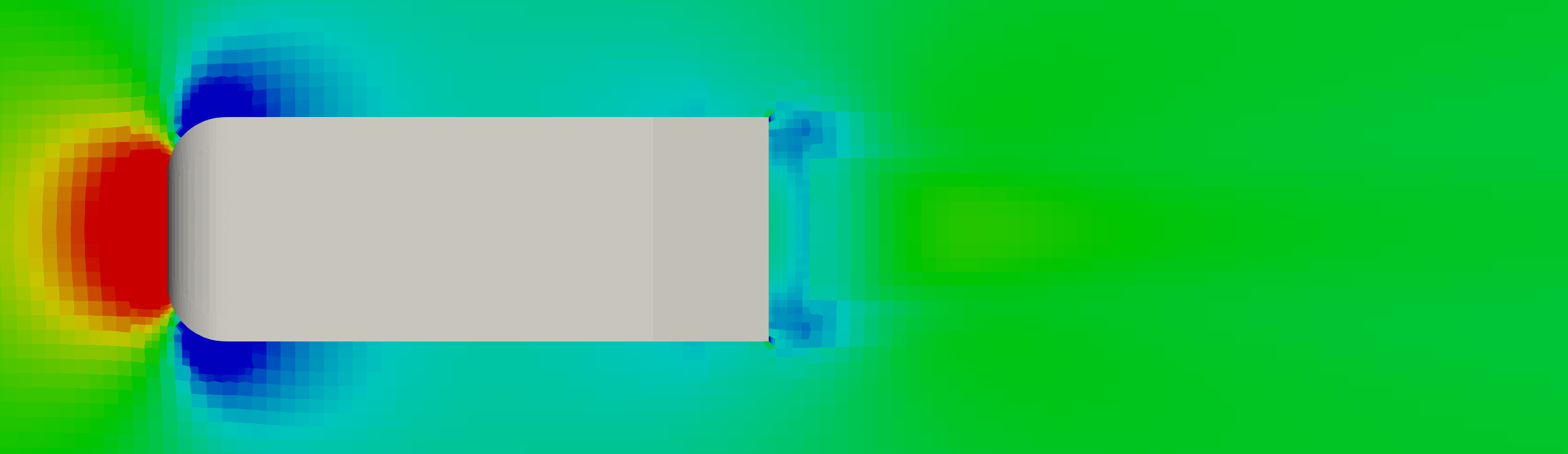}
    \end{subfigure}%
    
    \begin{subfigure}{0.36\textwidth}
    \centering
         \includegraphics[trim={0 0 0 0},clip,height=1.97cm]{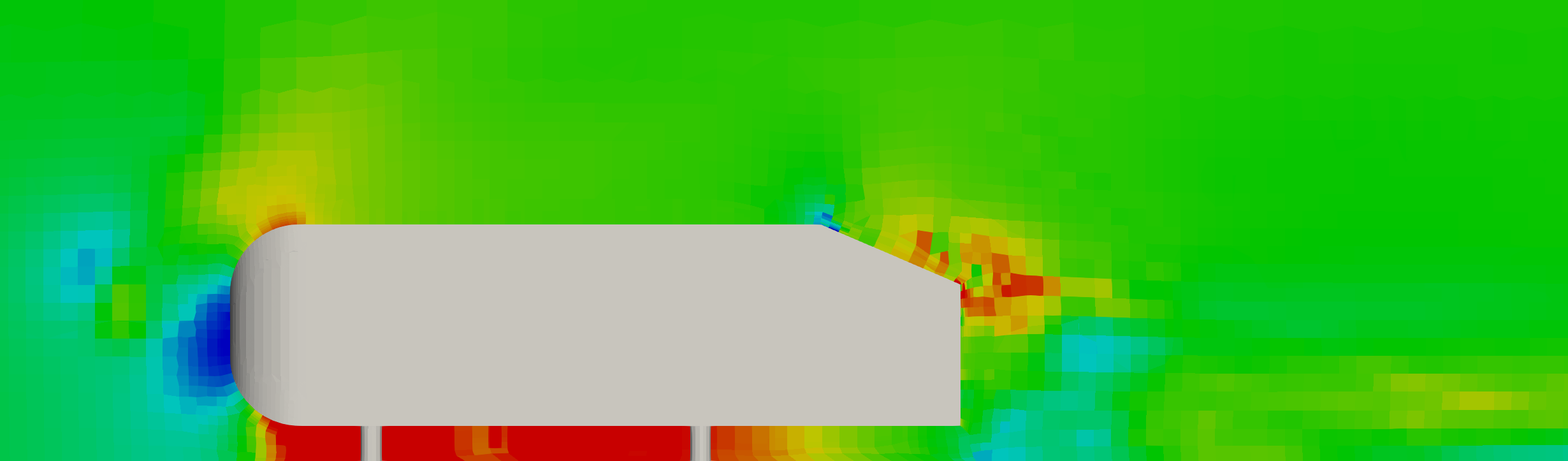}
    \end{subfigure}\hfill%
    \begin{subfigure}{0.36\textwidth}
    \centering
       \includegraphics[trim={0 0 0 0},clip,height=1.97cm]{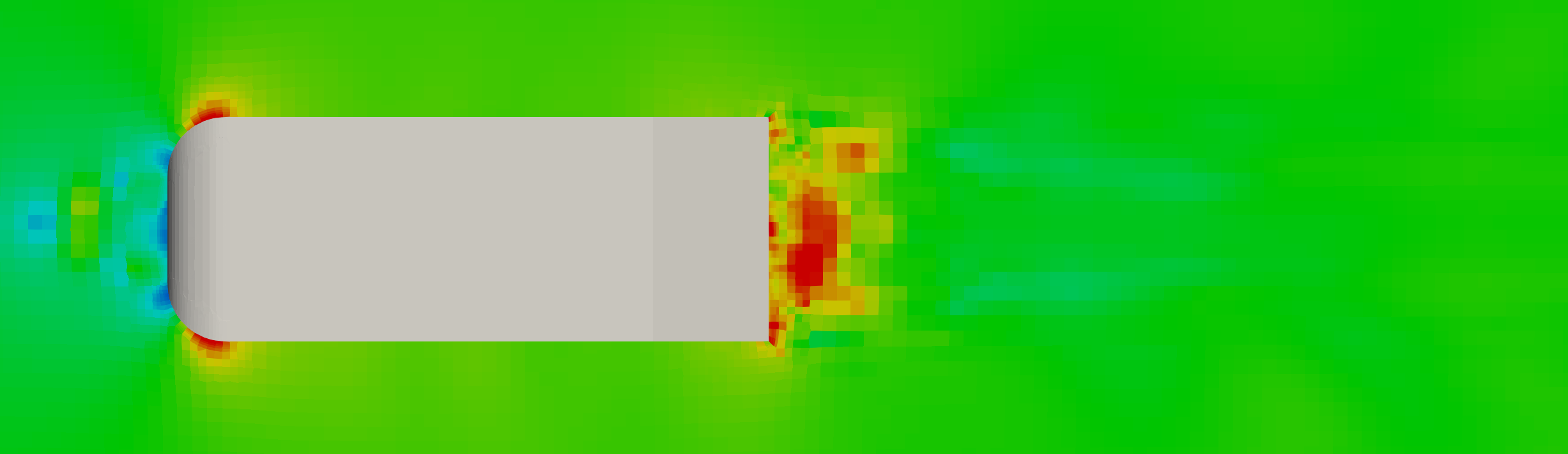}
    \end{subfigure}%

     \begin{subfigure}{0.99\textwidth}
    \centering
    \caption{Test sample with slant angle \SI{22.8}{\degree}}
    \end{subfigure}
      \vspace{0.7em}
    
     \begin{subfigure}{0.36\textwidth}
    \centering
     \begin{tikzpicture}
    \pgfplotscolorbardrawstandalone[ 
        colormap/bluered,
        colorbar sampled,
        colorbar horizontal,
        point meta min=-300,
        point meta max=400,
        colorbar style={
             font=\footnotesize,
            title=Pressure [\si{\pascal}],
            width=6cm,
            samples=256,}]
    \end{tikzpicture}
    \end{subfigure}
     \begin{subfigure}{0.36\textwidth}
    \centering
     \begin{tikzpicture}
    \pgfplotscolorbardrawstandalone[ 
        colormap/bluered,
        colorbar sampled,
        colorbar horizontal,
        point meta min=-5,
        point meta max=5,
        colorbar style={
             font=\footnotesize,
            title=Pressure Difference [\si{\pascal}],
            width=6cm,
            samples=256,}]
    \end{tikzpicture}
    \end{subfigure}
    
    \caption{Qualitative comparison for the pressure on the centerplane (left) and a slice $\SI{0.24}{\meter}$ above the street  (right): FOM results (top), ROM predictions (middle), and difference ($\text{ROM} - \text{FOM}$, bottom) for the test sample with lowest (a) and highest (b) slant angle.}
     \label{fig:rom_p_qualitative_comparison}
\end{figure}

Finally, \autoref{fig:rom_U_qualitative_comparison} and \autoref{fig:rom_p_qualitative_comparison}  compare the FOM and ROM fields qualitatively for velocity and pressure, respectively. We chose the test samples with the lowest and highest slant angle for this visual comparison.

For velocity and pressure, ROM and FOM results are in good agreement on both presented slices. In accordance with the quantitative results, for both fields, the errors for slant angle \SI{15.4}{\degree} are higher compared with those at \SI{22.8}{\degree}. 

As the parametrization alters the vehicle geometry exclusively at the rear end, the main flow field variations are expected to occur in the wake area of the vehicle; accordingly, for velocity, the highest ROM errors are visible in this region. Additionally, smaller regions at the top of the front end exhibit higher errors for both test samples. 

For the pressure, the regions of highest errors are scattered around the vehicle surface. Besides the wake region, in particular below the vehicle underbody high errors occur. The deficiencies of the pressure prediction of the ROM near the surface likely result in relatively high errors for the drag coefficients and is a topic of improvement for future work.




\section{Discussion}\label{sec:discussion}

In this paper we presented a new approach based on a technique that mixes up a classical projection-based method for what concerns both the momentum equation and the incompressibility constraint with a data-driven procedure for what regards the eddy viscosity closure. 

This choice revealed a wide applicability and flexibility since the turbulence model selected for the offline phase does not affect in any way the computations during the online phase. Moreover the reconstruction of the eddy viscosity field is very accurate as showed in \autoref{subsec:academicTC}.

The reduced SIMPLE algorithm we presented here in \autoref{subsec:rSimple}, taking advantage of the coupling between the accuracy of projection-based methods and the versatility of neural networks, showed to guarantee good approximations in widely different fluid dynamics test cases.
Moreover the idea of collecting converged fields together with middle iterations solutions ensures good convergence properties without showing relevant errors due to the physical information pollution of the modal basis functions, as explained in \autoref{subsec:ROM}.

Finally the choice of relying on an RBF approach for the mesh motion demonstrated to be effective while preserving a good shape of the modified mesh.

For what concerns the efficiency of the online phase of the problem, still some improvements are required and a natural forward step for this kind of applications would be the development of hyper reduction techniques for the reduced operators. This task could be also entrusted to neural networks approaches, trying to approximate the reduced operators by the evaluation, e.g., of an autoencoder. In any case the scope of this article was not focused on highly efficient hyper reduction techniques. Thus, even if in this procedure we are still relying on reconstructed full-dimension reduced order fields to assemble the equations, the results are in any case appreciable also in terms of time consuming. 


\vspace{6pt}

\section*{Acknowledgements}
We acknowledge the support by the European Commission H2020 ARIA (Accurate ROMs for Industrial Applications) project, by MIUR (Italian Ministry for Education University and Research) FARE-X-AROMA-CFD project and PRIN "Numerical Analysis for Full and Reduced Order Methods for Partial Differential Equations" (NA-FROM-PDEs) project, as well as the European Research Council Consolidator Grant  Advanced Reduced Order Methods with Applications in Computational Fluid Dynamics - GA 681447, H2020-ERC COG 2015 AROMA-CFD. Main computations in this work have been carried out by the usage of ITHACA-FV \cite{RoSta17}, a library maintained at SISSA mathLab, an implementation in OpenFOAM \cite{OpenFOAM} for reduced order modeling techniques; developers and contributors are acknowledged. 





\bibliographystyle{abbrv}
\bibliography{paperBibliography}

\begin{thebibliography}{10}

\bibitem{OpenFOAM}
{OpenFOAM} documentation website.
\newblock \url{https://openfoam.org/}. Accessed: 03/31/2021.

\bibitem{handbook1}
{\em {Model Order Reduction: Volume 1 System and Data-Driven Methods and
  Algorithms}}.
\newblock De Gruyter, 2020.

\bibitem{handbook2}
{\em {Model Order Reduction: Volume 2 Snapshot-Based Methods and Algorithms}}.
\newblock De Gruyter, 2020.

\bibitem{ahmed1984}
S.~Ahmed, G.~Ramm, and G.~Faltin.
\newblock Some salient features of the time-averaged ground vehicle wake.
\newblock In {\em SAE Technical Paper}. SAE International, 02 1984.

\bibitem{azaiez2021cure}
M.~Aza{\"\i}ez, T.~C. Rebollo, and S.~Rubino.
\newblock A cure for instabilities due to advection-dominance in pod solution
  to advection-diffusion-reaction equations.
\newblock {\em Journal of Computational Physics}, 425:109916, 2021.

\bibitem{ballarin2015supremizer}
F.~Ballarin, A.~Manzoni, A.~Quarteroni, and G.~Rozza.
\newblock Supremizer stabilization of pod--galerkin approximation of
  parametrized steady incompressible navier--stokes equations.
\newblock {\em International Journal for Numerical Methods in Engineering},
  102(5):1136--1161, 2015.

\bibitem{Benner2015}
P.~Benner, S.~Gugercin, and K.~Willcox.
\newblock {A Survey of Projection-Based Model Reduction Methods for Parametric
  Dynamical Systems}.
\newblock {\em {SIAM} Review}, 57(4):483--531, 2015.

\bibitem{bergmann2009enablers}
M.~Bergmann, C.-H. Bruneau, and A.~Iollo.
\newblock Enablers for robust pod models.
\newblock {\em Journal of Computational Physics}, 228(2):516--538, 2009.

\bibitem{bos2013radial}
F.~M. Bos, B.~W. van Oudheusden, and H.~Bijl.
\newblock Radial basis function based mesh deformation applied to simulation of
  flow around flapping wings.
\newblock {\em Computers \& Fluids}, 79:167--177, 2013.

\bibitem{Brunton2019}
S.~L. Brunton and J.~N. Kutz.
\newblock {\em {Data-Driven Science and Engineering}}.
\newblock Cambridge University Press, 2019.

\bibitem{busto2020pod}
S.~Busto, G.~Stabile, G.~Rozza, and M.~E. V{\'a}zquez-Cend{\'o}n.
\newblock Pod--galerkin reduced order methods for combined navier--stokes
  transport equations based on a hybrid fv-fe solver.
\newblock {\em Computers \& Mathematics with Applications}, 79(2):256--273,
  2020.

\bibitem{caiazzo2014numerical}
A.~Caiazzo, T.~Iliescu, V.~John, and S.~Schyschlowa.
\newblock A numerical investigation of velocity--pressure reduced order models
  for incompressible flows.
\newblock {\em Journal of Computational Physics}, 259:598--616, 2014.

\bibitem{chinesta2017model}
F.~Chinesta, A.~Huerta, G.~Rozza, and K.~Willcox.
\newblock Model reduction methods.
\newblock {\em Encyclopedia of Computational Mechanics Second Edition}, pages
  1--36, 2017.

\bibitem{de2007mesh}
A.~De~Boer, M.~Van~der Schoot, and H.~Bijl.
\newblock Mesh deformation based on radial basis function interpolation.
\newblock {\em Computers \& structures}, 85(11-14):784--795, 2007.

\bibitem{donea2003finite}
J.~Donea and A.~Huerta.
\newblock {\em Finite element methods for flow problems}.
\newblock John Wiley \& Sons, 2003.

\bibitem{drohmann2009reduced}
M.~Drohmann, B.~Haasdonk, and M.~Ohlberger.
\newblock Reduced basis method for finite volume approximation of evolution
  equations on parametrized geometries.
\newblock In {\em Proceedings of ALGORITMY}, volume 2008, pages 111--120, 2009.

\bibitem{dumon2011proper}
A.~Dumon, C.~Allery, and A.~Ammar.
\newblock Proper general decomposition (pgd) for the resolution of
  navier--stokes equations.
\newblock {\em Journal of Computational Physics}, 230(4):1387--1407, 2011.

\bibitem{GeorgakaStabileStarRozzaBluck2020}
S.~Georgaka, G.~Stabile, K.~Star, G.~Rozza, and M.~J. Bluck.
\newblock {A hybrid reduced order method for modelling turbulent heat transfer
  problems}.
\newblock {\em Computers \& Fluids}, 208:104615, 2020.

\bibitem{georgaka2020hybrid}
S.~Georgaka, G.~Stabile, K.~Star, G.~Rozza, and M.~J. Bluck.
\newblock A hybrid reduced order method for modelling turbulent heat transfer
  problems.
\newblock {\em Computers \& Fluids}, 208:104615, 2020.

\bibitem{goodfellow2016}
I.~Goodfellow, Y.~Bengio, and A.~Courville.
\newblock {\em Deep Learning}.
\newblock MIT Press, 2016.

\bibitem{hesthaven2016certified}
J.~S. Hesthaven, G.~Rozza, B.~Stamm, et~al.
\newblock {\em Certified reduced basis methods for parametrized partial
  differential equations}, volume 590.
\newblock Springer, 2016.

\bibitem{hijazi2020effort}
S.~Hijazi, S.~Ali, G.~Stabile, F.~Ballarin, and G.~Rozza.
\newblock The effort of increasing reynolds number in projection-based reduced
  order methods: from laminar to turbulent flows.
\newblock In {\em Numerical Methods for Flows}, pages 245--264. Springer, 2020.

\bibitem{HijaziStabileMolaRozza2020b}
S.~Hijazi, G.~Stabile, A.~Mola, and G.~Rozza.
\newblock {Data-Driven POD–Galerkin reduced order model for turbulent flows}.
\newblock {\em Journal of Computational Physics}, 416:109513, 2020.

\bibitem{hirsch2007numerical}
C.~Hirsch.
\newblock {\em Numerical computation of internal and external flows: The
  fundamentals of computational fluid dynamics}.
\newblock Elsevier, 2007.

\bibitem{iapichino2014reduced}
L.~Iapichino, A.~Quarteroni, G.~Rozza, and S.~Volkwein.
\newblock Reduced basis method for the stokes equations in decomposable
  parametrized domains using greedy optimization.
\newblock In {\em European Consortium for Mathematics in Industry}, pages
  647--654. Springer, 2014.

\bibitem{iollo2000stability}
A.~Iollo, S.~Lanteri, and J.-A. D{\'e}sid{\'e}ri.
\newblock Stability properties of pod--galerkin approximations for the
  compressible navier--stokes equations.
\newblock {\em Theoretical and Computational Fluid Dynamics}, 13(6):377--396,
  2000.

\bibitem{jasak1996error}
H.~Jasak.
\newblock Error analysis and estimation for the finite volume method with
  applications to fluid flows.
\newblock 1996.

\bibitem{kim2020efficient}
Y.~Kim, Y.~Choi, D.~Widemann, and T.~Zohdi.
\newblock Efficient nonlinear manifold reduced order model.
\newblock {\em arXiv preprint arXiv:2011.07727}, 2020.

\bibitem{kingma2014}
D.~Kingma and J.~Ba.
\newblock Adam: A method for stochastic optimization, 2014.

\bibitem{lee2020model}
K.~Lee and K.~T. Carlberg.
\newblock Model reduction of dynamical systems on nonlinear manifolds using
  deep convolutional autoencoders.
\newblock {\em Journal of Computational Physics}, 404:108973, 2020.

\bibitem{moukalled2016finite}
F.~Moukalled, L.~Mangani, M.~Darwish, et~al.
\newblock {\em The finite volume method in computational fluid dynamics},
  volume 113.
\newblock Springer, 2016.

\bibitem{paszke2019}
A.~Paszke, S.~Gross, F.~Massa, A.~Lerer, J.~Bradbury, G.~Chanan, T.~Killeen,
  Z.~Lin, N.~Gimelshein, L.~Antiga, A.~Desmaison, A.~Kopf, E.~Yang, Z.~DeVito,
  M.~Raison, A.~Tejani, S.~Chilamkurthy, B.~Steiner, L.~Fang, J.~Bai, and
  S.~Chintala.
\newblock Pytorch: An imperative style, high-performance deep learning library.
\newblock In H.~Wallach, H.~Larochelle, A.~Beygelzimer, F.~d\textquotesingle
  Alch\'{e}-Buc, E.~Fox, and R.~Garnett, editors, {\em Advances in Neural
  Information Processing Systems 32}, pages 8024--8035. Curran Associates,
  Inc., 2019.

\bibitem{rumelhart1986}
D.~E. Rumelhart, G.~E. Hinton, and R.~J. Williams.
\newblock {Learning Representations by Back-Propagating Errors}.
\newblock {\em Nature}, 323(6088):533--536, 1986.

\bibitem{salmoiraghi2018free}
F.~Salmoiraghi, A.~Scardigli, H.~Telib, and G.~Rozza.
\newblock Free-form deformation, mesh morphing and reduced-order methods:
  enablers for efficient aerodynamic shape optimisation.
\newblock {\em International Journal of Computational Fluid Dynamics},
  32(4-5):233--247, 2018.

\bibitem{stabile2017advances}
G.~Stabile, S.~Hijazi, A.~Mola, S.~Lorenzi, and G.~Rozza.
\newblock Advances in reduced order modelling for cfd: vortex shedding around a
  circular cylinder using a pod-galerkin method.
\newblock {\em arXiv preprint arXiv:1701.03424}, 945, 2017.

\bibitem{RoSta17}
G.~Stabile and G.~Rozza.
\newblock {ITHACA-FV - In real Time Highly Advanced Computational Applications
  for Finite Volumes}.
\newblock \url{http://www.mathlab.sissa.it/ithaca-fv}. Accessed: 03/31/2021.

\bibitem{stabile2018finite}
G.~Stabile and G.~Rozza.
\newblock Finite volume pod-galerkin stabilised reduced order methods for the
  parametrised incompressible navier--stokes equations.
\newblock {\em Computers \& Fluids}, 173:273--284, 2018.

\bibitem{stabile2020efficient}
G.~Stabile, M.~Zancanaro, and G.~Rozza.
\newblock {Efficient Geometrical parametrization for finite-volume based
  reduced order methods}.
\newblock {\em International Journal for Numerical Methods in Engineering},
  121(12):2655--2682, 2020.

\bibitem{tsiolakis2020nonintrusive}
V.~Tsiolakis, M.~Giacomini, R.~Sevilla, C.~Othmer, and A.~Huerta.
\newblock Nonintrusive proper generalised decomposition for parametrised
  incompressible flow problems in openfoam.
\newblock {\em Computer physics communications}, 249:107013, 2020.

\bibitem{wang2012proper}
Z.~Wang, I.~Akhtar, J.~Borggaard, and T.~Iliescu.
\newblock Proper orthogonal decomposition closure models for turbulent flows: a
  numerical comparison.
\newblock {\em Computer Methods in Applied Mechanics and Engineering},
  237:10--26, 2012.

\bibitem{wilcox1998turbulence}
D.~C. Wilcox et~al.
\newblock {\em Turbulence modeling for CFD}, volume~2.
\newblock DCW industries La Canada, CA, 1998.

\end{thebibliography}

\end{document}